\documentclass[twocolumn]{aastex7}
\usepackage{amsmath}
\usepackage{nccmath}
\usepackage{newtxtext,newtxmath}
\usepackage{amssymb}
\usepackage{latexsym}
\usepackage{graphicx}
\usepackage{longtable}
\usepackage[english]{babel}
\usepackage[utf8x]{inputenc}
\usepackage{listings}
\usepackage{subfigure}
\usepackage{float}
\usepackage{url}
\usepackage[colorinlistoftodos]{todonotes}
\usepackage{hyperref}

\usepackage{enumitem}
\usepackage{booktabs}

\usepackage{natbib}
\setcitestyle{aas}

\begin{document}

\title{The Photometric Analysis of the Environment Around Two Dusty Star-Forming Galaxies at $z \sim 2$}
\author[0009-0007-8574-5720]{Joe Bhangal}
\affiliation{Department of Physics \& Astronomy, University of British Columbia, 6224 Agricultural Road, Vancouver, BC V6T 1Z1, Canada \\}
\email{joey.bhangal@ubc.ca}

\author[0000-0003-2475-124X]{Allison W. S. Man}
\affiliation{Department of Physics \& Astronomy, University of British Columbia, 6224 Agricultural Road, Vancouver, BC V6T 1Z1, Canada \\}
\email{aman@phas.ubc.ca}

\author[0000-0002-5268-2221]{Tom J. L. C. Bakx}
\affiliation{Department of Earth and Space Sciences, Chalmers University of Technology, Onsala Observatory, SE-439 94 Onsala, Sweden\\}
\email{tom.bakx@chalmers.se}

\author[0000-0001-5341-2162]{Darko Donevski}
\affiliation{National Center for Nuclear Research, Pasteura 7, 02-093 Warsaw, Poland\\}
\affiliation{SISSA, Via Bonomea 265, 34136, Trieste, Italy\\}
\email{darko.donevski@ncbj.gov.pl}

\author[0000-0003-2027-8221]{Pierre Cox}
\affiliation{Sorbonne Université, UPMC Université Paris 6 and CNRS, UMR 7095, Institut d’Astrophysique de Paris, 98bis Boulevard Arago, F-75014 Paris, France\\}
\email{cox@iap.fr}

\author[0000-0001-7147-3575]{Helmut Dannerbauer}
\affiliation{Instituto de Astrof\'isica de Canarias, C/ V\'ia L\'actea s/n, 38205 La Laguna, Tenerife, Spain}
\affiliation{Departamento de Astrof\'isica, Universidad de La Laguna, 38205 La Laguna, Tenerife, Spain}

\email{helmut@iac.es}

\author[0000-0002-0517-7943]{Stephen Serjeant}
\affiliation{School of Physical Sciences, The Open University, Milton Keynes,
MK7 6AA, UK\\}
\email{stephen.serjeant@open.ac.uk}

\author[0000-0001-8083-5814]{Masato Hagimoto}
\affiliation{Department of Physics, Graduate School of Science, Nagoya University, Nagoya, Aichi 464-8602, Japan\\}
\email{hagimoto@a.phys.nagoya-u.ac.jp}

\author{Pluto Jiang}
\affiliation{Department of Physics \& Astronomy, University of Waterloo, 200 University Ave W, Waterloo, ON N2L 3G1, Canada \\}
\affiliation{Department of Physics \& Astronomy, University of British Columbia, 6224 Agricultural Road, Vancouver, BC V6T 1Z1, Canada \\}
\email{pluto.jiang@uwaterloo.ca}

\author{Wenxiao Liu}
\affiliation{Department of Physics \& Astronomy, University of British Columbia, 6224 Agricultural Road, Vancouver, BC V6T 1Z1, Canada \\}
\email{lwx02121999@gmail.com}


\begin{abstract}

Studying the environments of dusty star-forming galaxies (DSFGs) provides insight into whether these luminous systems are reliable signposts of large-scale overdensities. Evidence suggests that individual DSFGs can trace overdense environments, although this association may not be universal. To test this, we investigate the environments surrounding two luminous, gravitationally-lensed DSFGs (SDP.17b at $z_\text{spec} = 2.3049$ and HELMS-55 at $z_\text{spec} = 2.2834$). Using Gemini South Flamingos-2 (F2) $K_s$-band imaging together with ancillary Subaru Hyper Suprime-Cam and Hubble Space Telescope multi-band photometry, we obtain photometric redshifts, $z_\text{phot}$, as well as star formation rates and stellar mass estimates for companion galaxies of the DSFGs. At least $5\pm2$ and $15\pm3$ companion galaxies exist with consistent $z_\text{phot}$  ($dz \leq 0.2$) within a projected separation of 5.5 cMpc of SDP.17b and HELMS-55, respectively. These correspond to galaxy overdensities of $\delta = 0.1 \pm 0.2$ and ${\delta} =1.0 \pm 0.3$, with significances of $(0.2 \pm 0.4)\sigma$ and $(2.2 \pm 0.6) \sigma$, respectively. On the $M_{\rm H_2}$-overdensity-significance plane, HELMS-55 may follow the positive correlation between the gas mass and the overdensity significance, while SDP.17b lies well above the relation despite its large gas reservoir, making it a potential outlier. Based on this study of two DSFGs, our photometric analysis suggests that DSFGs can trace the outskirts of protoclusters or associated large-scale structures. However, our small sample prevents firm conclusions about their ability to pinpoint dense cluster cores. Future multi-object spectroscopic observations are required to confirm the membership and star formation properties of the companion galaxies.

\end{abstract}



\section{Introduction}\label{sec:intro}

As the most massive gravitationally bound systems in the Universe, galaxy clusters play a central role in tracing the growth of large-scale structure. Studying their progenitors, known as protoclusters, can help provide a better understanding of the assembly of clusters and cluster galaxies \citep{Overzier2016, Alberts_2022}. There are varying definitions of protoclusters in the literature, however, they are commonly defined as overdensities of galaxies or dark matter haloes at high redshift that will eventually collapse into a massive, virialized galaxy cluster \citep[halo mass $\geq 10^{14}\, M_\odot$ at $z \geq 0$]{Overzier2016}. Protoclusters are expected to have dark matter halos ranging in radius from $\sim$1--10 comoving Mpc (cMpc) based on the Millennium cosmological $N$-body simulation \citep{Springel_2005} and semi-analytic models \citep{Chiang_2017}; however, there is no unique definition of protocluster size, which makes it challenging to determine whether different selection methods agree and are equally accurate \citep{Muldrew_2015, Lovell_2018, Lim_2024}.

Observationally, no consensus exists on how protocluster environments affect galaxy evolution in terms of star formation rates (SFRs) and gas content. Protocluster galaxies are found to have boosted \citep{Shimakawa_2018, Monson_2021}, suppressed \citep{Tran_2015}, or similar SFRs \citep{Kiyota_2024} compared to field galaxies. They are also reported to be more gas-rich than field galaxies \citep{Noble_2017, Tadaki_2019} or similar in gas content \citep{Castignani_2020, Aoyama_2022}. The diversity of observational results requires a combined perspective from models and well-defined observations to better understand the role of the environment in galaxy evolution.

Cosmological simulations have predicted that protoclusters contribute significantly to the stellar mass budget of the Universe (e.g., \citealt{Bassini_2020}). However, many struggle to reproduce key observables, particularly the high star formation rates (SFRs) seen in protocluster cores at $z > 2$, in part due to the lack of realistic interstellar medium (ISM) modeling \citep{Lim_2021}. This highlights the need for further observational studies of high-redshift protoclusters to refine these models.

Both observational \citep{Capak_2011, Dannerbauer_2014, Umehata2015, Miller_2018, Oteo_2018, Harikane_2019, Long_2020, Calvi_2023} and theoretical studies \citep{Araya_Araya_2024} suggest that dusty star-forming galaxies (DSFGs; also known as bright submillimeter galaxies; SMGs) may be strongly associated with overdense environments at $2<z<6$. DSFGs are often found in some of the most massive dark matter haloes at a given epoch \citep{Marrone_2018, Garcia_Vergara_2020, Stach_2021, Araya_Araya_2024}, making them promising signposts for protoclusters and massive halo environments \citep{Wilkinson_2017, Donevski_2018, Miller_2018}. However, the reliability of DSFGs as protocluster tracers remains debated. Several studies argue that DSFGs, especially bright ones, are incomplete tracers of overdensities due to their rarity and short starburst duty cycle \citep{Chapman_2009, Chapman_2015, Miller_2015, Casey_2016, Alvarez_2021, Gao_2022}. For instance, \citet{Gao_2022} found that $>96\%$ of their protocluster candidates do not host bright DSFGs, and hypothesized that this is due to the short-lived enhanced starbursting phase of DSFGs. Even one of the most luminous (unlensed) DSFGs, HS1700.850.1 at $z = 2.82$, was found to reside in an underdense galaxy environment by \citet{Chapman_2015}. In contrast, \citet{Calvi_2023} systematically found that 11 out of 12 of DSFGs in their study ($92\pm 8\%$) reside in galaxy overdensities, suggesting that at least some populations of DSFGs can be effective tracers of protocluster environments. Some studies find a diversity of environments, with \citet{Cornish_2024} reporting DSFGs located in significant and moderate overdensities and field-like regions. This tension highlights the importance of better understanding which selection of DSFGs (e.g., bright vs. faint, lensed vs. unlensed) is most predictive of large-scale structure.

The identification of overdensities and protocluster environments around DSFGs has relied on a variety of methods, each introducing selection biases that shape our interpretation of DSFGs as tracers. For instance, identifying overdensity candidates using deep millimeter imaging surveys with the South Pole Telescope (SPT) followed by ALMA spectral scans (e.g., \citealt{Miller_2018}) tends to favor strongly lensed or intrinsically luminous systems due to sensitivity limitations. Similarly, single-dish observations with instruments like SCUBA-2 are biased toward the brightest dusty galaxies (e.g., \citealt{Chapman_2015, Wilkinson_2017}). Although ALMA follow-up can resolve blended sources into multiple components (e.g., \citealt{Oteo_2018}), its narrow field of view (FOV) may miss fainter, spatially offset companions. Other approaches, such as identifying overdensities via Ly$\alpha$ or H$\alpha$ emission with ALMA or Keck (e.g., \citealt{Umehata2015, Harikane_2019}), are biased against dusty or Ly$\alpha$-faint galaxies, potentially missing obscured members. Each of these methods carries its own selection biases, influencing how we interpret DSFG environments. Although spectroscopic confirmation provides robust redshift information, it is observationally expensive and limited to brighter sources. In contrast, photometric redshifts (e.g., \citealt{Calvi_2023}) and submillimeter (submm) surveys enable larger samples but suffer from redshift uncertainties. 

Compared to other tracers of galaxy overdensities, such as high-redshift radio galaxies (HzRGs), DSFGs are at least 10 times more common \citep{Reuland_2003, Calvi_2023}, and the negative K-correction allows them to be detected independent of redshift at a fixed luminosity \citep{Blain_2002}. Therefore, this paper aims to test whether luminous DSFGs have overdense environments characteristic of protoclusters. To search for potential structures around luminous DSFGs, we survey a wide area around the central source, sufficiently deep at a NIR wavelength (a tracer of stellar mass in rest-frame optical), hence the need for deep Gemini $K_s$-band data. 

In this paper, we present new Gemini South Flamingos-2 $K_s$-band imaging, complemented by ancillary Subaru Hyper Suprime-Cam (HSC) optical and Hubble Space Telescope (HST) NIR imaging data. We test whether DSFGs are tracers of overdense regions by quantifying the overdensity of the environments of two DSFGs, namely: SDP.17b at $z_\text{spec}=2.3049$ and HELMS-55 at $z_\text{spec}=2.2834$. This paper is organized as follows. Section \ref{sec:target_selec_and_data} describes the target selection and data set used. We describe the source detection and photometry, photometric redshift fits, overdensity analysis, and estimating the properties of galaxies in Section \ref{sec:Methods}. We conclude the paper with a discussion of our results and future plans in Sections \ref{sec:results_and_implications} and \ref{sec:conclusions}. All magnitudes are expressed as AB magnitudes for this paper \citep{Oke_1974}. Throughout this paper, we adopt a spatially flat $\Lambda$CDM cosmology with $H_0$ = 67.66 $\text{km}\, \text{s}^{-1} \, \text{Mpc}^{-1}$ and $\Omega_\text{M}$ = 0.30966 (\hspace{-0.06cm}\citealt{Planck_2018_2020_paper}), and a \citet{Chabrier_2003} initial mass function (IMF) is assumed.

\section{Target Selection and Data}\label{sec:target_selec_and_data}

\subsection{SPOTLESS}\label{subsec:spotless}

\begin{figure}
    \centering
    \includegraphics[width=1\linewidth]{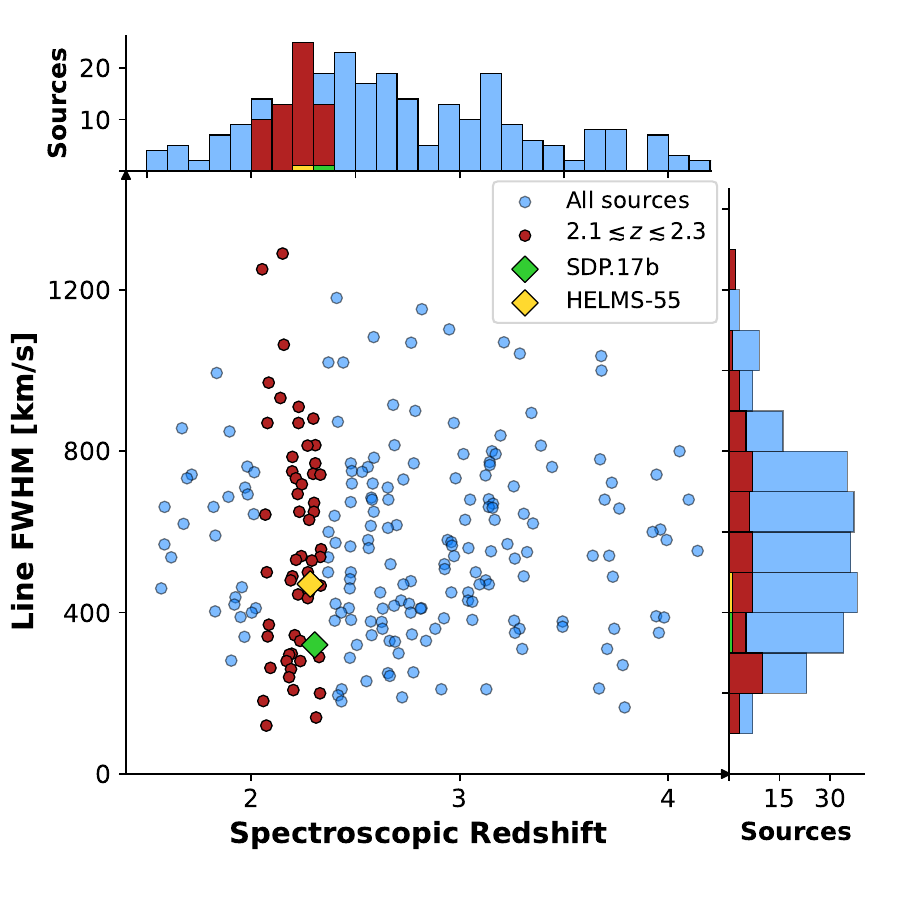}
    \caption{A subset of the SPOTLESS sample of \textit{Herschel} DSFGs with available line FWHM measurements from the literature. Redshifts are determined via targeted, mostly interferometric, submillimeter CO line observations. The $y$-axis shows the FWHM of the line, serving as a proxy for the galaxy’s internal kinematics. Red markers indicate galaxies at $2.1 \lesssim z \lesssim 2.3$, where prominent spectral features fall within the $K$-band for MOS follow-up. Green and yellow markers highlight the two DSFGs studied in this paper: SDP.17b and HELMS-55, respectively. Blue markers denote the rest of the SPOTLESS sample. This plot is zoomed in to redshifts $1.5 \lesssim z \lesssim 4$ and FWHM $\leq 1450$ km s$^{-1}$ to highlight the bulk population of the sample.}
    \label{fig:SPOTLESS_sample}
\end{figure}

We analyze the environments around two DSFGs: SDP.17b ($z_\text{spec} = 2.3049 \pm 0.0006$; \citealt{Omont_2011}) and HELMS-55  ($z_\text{spec} = 2.2834 \pm 0.0002$; \citealt{Cox_2023}). These two DSFGs were selected from the \textbf{S}earching for \textbf{P}rotoclusters through \textbf{O}bservations \textbf{T}racing \textbf{L}uminous \textbf{E}xtragalactic \textbf{S}ources \textbf{S}urvey (SPOTLESS), a collection of dusty galaxies drawn from \textit{Herschel} surveys. SPOTLESS is based on an extensive decade-long spectroscopic redshift campaign targeting $\sim$300 \textit{Herschel}-selected DSFGs over 1000 $\text{deg}^2$ (\citealt{Swinbank_2010, Fu_2013, Negrello_2017, Yang_2017, Wu_2019, Neri_2020, Urquhart_2022, Cox_2023} \& references therein). The SPOTLESS sample is shown in Figure \ref{fig:SPOTLESS_sample} and is available in our GitHub repository\footnote{\label{SPOTLESS_footnote}\url{https://github.com/Joe-Bhangal/SPOTLESS}} and on Zenodo \citep{SPOTLESS_Zendo}. The spectroscopic redshifts of this subset of {\it Herschel}-selected DSFGs enable the characterization of galaxies in their immediate environment. SPOTLESS targets are selected at $z \sim 2$ such that H$\alpha$ emission falls within the $K$-band, making them suitable for observation with multi-object spectrographs (MOS) and/or narrowband filters. 

The sources in the SPOTLESS sample show a wide distribution of CO line widths, extending to $\sim 1200$ km s$^{-1}$ for the broadest lines, with a median value of the full width at half maximum (FWHM) of $\sim 600$ km s$^{-1}$, measured across multiple transitions (e.g., CO(3–2), CO(4–3), CO(2–1)). SDP.17b and HELMS-55 have linewidths of $320\pm10$ [CO(4-3)] and $471\pm29$ km\,s$^{-1}$ [CO(3-2)], respectively, placing them toward the lower end of the SPOTLESS sample, suggesting moderate dynamical masses or low gas turbulence. In the apparent CO luminosity ($L'_{\rm CO}$) vs FWHM relation of DSFGs \citep{Greve_2005, Bothwell_2013, Jin_2021, Cox_2023}, both sources fall well above the relation due to their high apparent CO luminosities, consistent with the effects of gravitational lensing.

This pilot study focuses on the NIR continuum imaging of two bright DSFGs. SDP.17b was originally detected in the \textit{Herschel}-ATLAS Science Demonstration Phase (SDP; \citealt{Eales_Hershel_Atlas_2010}). HELMS-55 originates from the \textit{Herschel} Multi-tiered Extragalactic Survey (HerMES) Large Mode Survey (HELMS; \citealt{Asboth_2016}). Both SDP.17b and HELMS-55 are galaxy-galaxy gravitational lens systems \citep{Negrello_2010, Negrello_2014, Nayyeri_2016, Borsato_2024}. SDP.17b is lensed by a foreground galaxy at a spectroscopic redshift of $z_\text{spec} = 0.9435 \pm 0.0009$ (SDP.17a; \citealt{Negrello_2014}), shown in magenta in the RGB cutout in Figure \ref{fig:SDP17_HELMS-55_data}. HELMS-55 is included as a lens candidate in \citet{Nayyeri_2016}, and subsequent imaging with HST \citep{Borsato_2024} and HSC (Bakx et al., in prep.) confirms that it is gravitationally lensed by a bright foreground galaxy.

\subsection{Gemini South Flamingos-2 Ks-band Imaging}\label{subsec:Ks_band_imaging}

We present the first results of the SPOTLESS-Gemini program, in which we use the Flamingos-2 (F2) instrument to obtain wide-field $K_s$-band imaging towards SDP.17b and HELMS-55. F2 is an imager with a circular FOV spanning 6.2 arcmin in diameter, and it also operates as a long-slit and multi-object spectrograph \citep{Flamingos_2_2004, Flaminogos_2_2022}. SDP.17b and HELMS-55 were chosen as pilot targets for the SPOTLESS-Gemini program since they have ancillary optical observations to enable photometric redshift estimates and thus to test the overdensity hypothesis. A summary of the photometric data for each target environment is shown in Figure \ref{fig:SDP17_HELMS-55_data}. The environment around SDP.17b was observed in the $K_s$-band for $9587\,\mathrm{s}$ from October 2022 to January 2023 (program ID GS-2022B-Q-312; PI: Allison Man); however, $6644\,\mathrm{s}$ of the data was contaminated by artifacts. Therefore, follow-up $K_s$-band observations were conducted for an additional $9072\,\mathrm{s}$ in April 2025 (program ID GS-2025A-Q-308; PI: Joe Bhangal). The environment around HELMS-55 was observed in the $K_s$-band for $9782\,\mathrm{s}$ in September 2023 (program ID GS-2023B-Q-316; PI: Allison Man). Appendix \ref{subsec:observation_log} contains the full observation log for all programs. Observations of the SDP.17b and HELMS-55 environments were conducted under image quality conditions ranging from 70–85\% (good to poor) and cloud cover between 50–70\% (clear to cirrus). Table \ref{tab:photometric_data} provides information about the quality of the observations used for this analysis, including the point-spread function (PSF) FWHM and the $3\sigma$ point-source sensitivity within a $2''$ diameter aperture.

The $K_s$-band imaging data was processed using \href{https://www.gemini.edu/observing/phase-iii/reducing-data/dragons-data-reduction-software}{\texttt{DRAGONS}} (Data Reduction for Astronomy from Gemini Observatory North and South; \citealt{DRAGONS_2023, Zendo_dragons}). The calibrated $K_s$-band images have a pixel scale of $\sim$0.179 arcsec/pixel. Following the observatory data calibration advice, \textit{H}-band flat frames were used to create a master flat to correct for the transmission across the FOV, as the $K_s$-band flat frames are affected by the variable thermal emission from the telescope\footnote{\url{https://gemini-iraf-flamingos-2-cookbook.readthedocs.io/en/latest/Processing.html}}. A customized version of the \texttt{DRAGONS} \texttt{ultradeep} calibration recipe was used to calibrate the $K_s$-band images. Two modifications were made to the default \texttt{DRAGONS} \texttt{ultradeep} calibration recipe that improved the calibration. To define the statistics for scaling the sky frame to the science frames, one specific region for each target environment was used instead of the entire F2 $K_s$-band images. These regions were manually chosen to give a more accurate level for the sky without artifacts or bright stars. We also extended the time constraint for defining sky frames to include the full length of the dither pattern. Artifacts were discovered in the SDP.17b environment's 2022/2023 $K_s$-band imaging. At the moment, there is no method to correct them, so the contaminated data ($6644\,\mathrm{s}$) was discarded, and new observations were conducted.

The PSF FWHMs of the $K_s$-band images were determined by fitting 2D Gaussians to stacked cutouts of known stars in the image (see Section \ref{subsec:detection_and_photometry} for more details). \href{https://astrometry.net/}{\texttt{Astrometry.net}} \citep{Astrometry_Net_2010} was used to correct for astrometric offsets to the J2000 reference frame for each image.

\begin{table}[t]
    \centering
    \begin{tabular}{ccccc}
        \toprule
        \textbf{Filter} & \textbf{PSF}& \textbf{$3\sigma$} & \textbf{Zeropoint} & \textbf{Usable} \\
         & \textbf{FWHM} & \textbf{Depth} & & \textbf{Area} \\
         & [$''$] & [AB mag]& [AB mag]& [$\text{arcmin}^2$] \\
        \midrule
        \multicolumn{5}{c}{\textbf{SDP.17b Environment}} \\
        \midrule
        \textit{g}   & 0.77 & 25.66 & $27.0206 \pm 0.0006$ & $>$44 \\
        \textit{r}   & 1.09 & 24.89 & $27.0439 \pm 0.0007$ & $''$\\
        \textit{i}   & 0.79 & 24.67 & $27.015 \pm 0.002$   & $''$\\
        \textit{z}   & 0.74 & 24.32 & $27.014 \pm 0.003$   & $''$\\
        \textit{y}   & 0.78 & 23.78 & $27.028 \pm 0.001$   & $''$\\
        F110W        & 0.23 & 24.58 & $26.82^a$            & 4.75 \\
        F160W        & 0.28 & 24.27 & $25.94^a$            & 4.75 \\
        $K_s$        & 0.70 & 22.91 & $30.87 \pm 0.01$     & 30.19 \\
        \midrule
        \multicolumn{5}{c}{\textbf{HELMS-55 Environment}} \\
        \midrule
        \textit{g}   & 0.52 & 25.30 & $27.0885 \pm 0.0004$ & $>$44 \\
        \textit{r}   & 0.50 & 24.92 & $27.0275 \pm 0.0005$ & $''$\\
        \textit{i}   & 0.61 & 24.71 & $27.027 \pm 0.004$   & $''$\\
        \textit{z}   & 0.74 & 24.35 & $27.0108 \pm 0.0007$ & $''$\\
        \textit{y}   & 0.76 & 23.83 & $27.022 \pm 0.001$   & $''$\\
        F110W        & 0.23 & 24.10 & $26.82^a$            & 4.74 \\
        $K_s$        & 0.73 & 23.10 & $30.43 \pm 0.02$     & 31.03 \\
        \bottomrule
    \end{tabular}
    \caption{A summary of the photometric data used for the SDP.17b and HELMS-55 target environments. The HSC \textit{grizy} imaging covers beyond the F2 footprint, and the HST F110W and F160W imaging covers a fraction of the footprint of the other bands. In all rows, we show the native PSF FWHM, before resampling to match the pixel scale of the $K_s$-band images. Each depth value corresponds to the $3\sigma$ point-source sensitivity within a $2''$ diameter aperture. Note HELMS-55 does not have HST F160W band coverage.  
    (a) The AB zeropoints of the HST bands were computed using their respective science image's header information.}
    \label{tab:photometric_data}
\end{table}

\subsection{Ancillary Data}\label{subsec:ancillary_data}

Ancillary optical \textit{grizy} band images were obtained from the Subaru HSC Public Data Release 3 (PDR3; \citealt{Aihara_2022_HSC_PDR3}) via the HSC query service\footnote{\url{https://hsc-release.mtk.nao.ac.jp/datasearch/}}, along with their respective weight maps. These optical images cover the full $K_s$-band F2 footprint for both target environments and have a pixel scale of $\sim$0.168 arcsec/pixel. The images are already astrometrically calibrated in the HSC PDR3 pipeline, using GAIA Data Release 1 (\hspace{-0.06cm}\citealt{GAIA_DR1}). 

\begin{figure*}
    \centering
    \includegraphics[width=\linewidth]{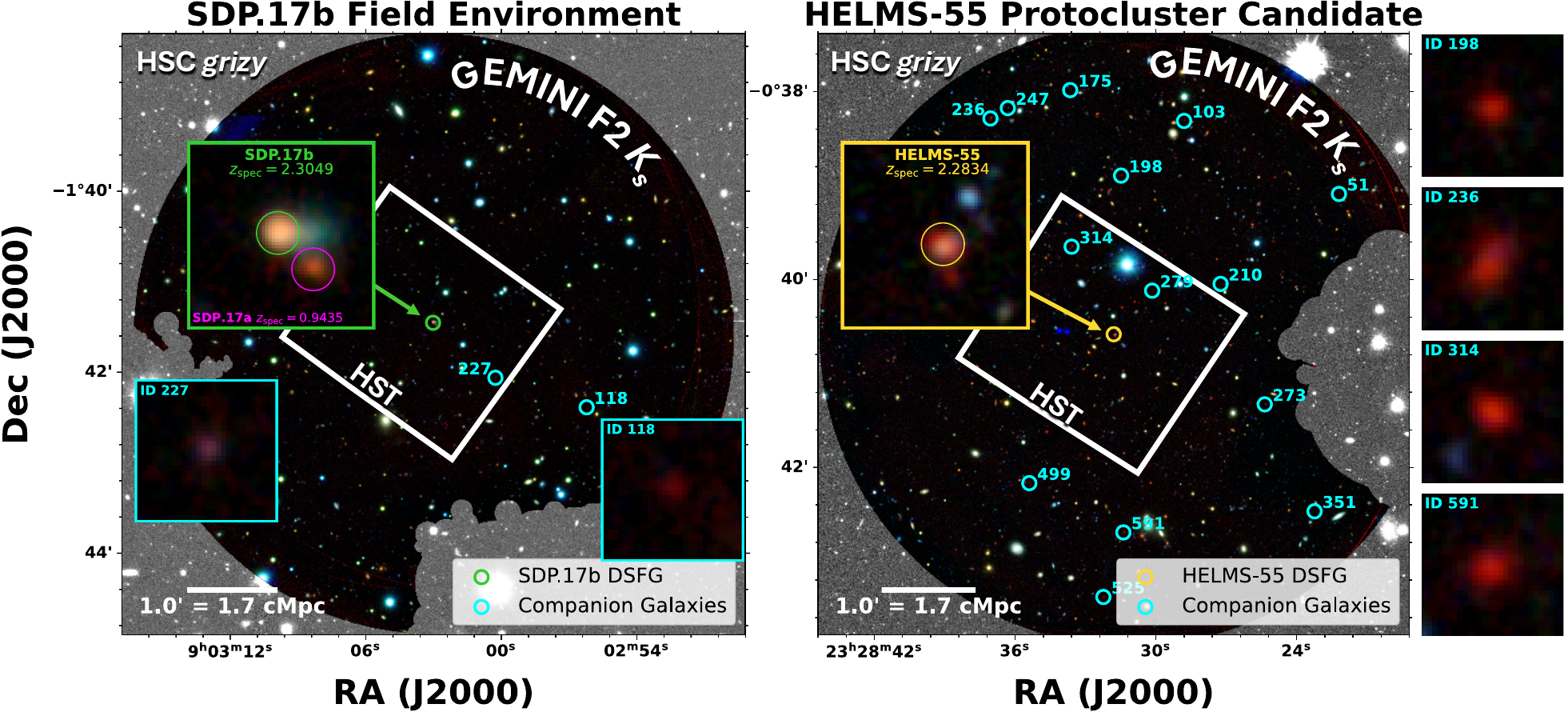}
    \caption{Summary of the photometric data for each target environment, including companion galaxies associated with the central DSFGs. Gemini South F2 $K_s$-band footprints are shown as RGB images: $K_s$ (red), $i$ (green), and $g$ (blue). Gaps in the F2 circular FOV arise from masking the OIWFS probe shadows during observations. The HSC optical \textit{grizy} footprint (grayscale) extends beyond the F2 coverage, and HST F110W/F160W footprints are shown as white $\sim 2.05' \times 2.27'$ rectangles. Companion galaxies are marked with cyan circles, while the central DSFG is highlighted in green for SDP.17b and yellow for HELMS-55. Example companion galaxies are displayed in $5'' \times 5''$ RGB cutouts, with their object ID indicated. Central DSFGs are shown in larger $10'' \times 10''$ cutouts. SDP.17a is circled in magenta in SDP.17b's cutout.}
    \label{fig:SDP17_HELMS-55_data}
\end{figure*}

HST NIR Wide Field Camera-3 (WFC3) imaging was also used, however, these images only cover the central portion of the F2 footprint ($\sim 2.05' \times 2.27'$ rectangle). The images were retrieved from the Mikulski Archive for Space Telescopes (MAST)\footnote{\url{https://mast.stsci.edu/search/ui/\#/hst}} and have a native pixel scale of $\sim 0.128$ arcsec/pixel. The HST images were calibrated by the Space Telescope Institute (STScI) using the standard WFC3 calibration pipeline\footnote{\url{https://www.stsci.edu/hst/instrumentation/wfc3/software-tools/pipeline}}, which provided their respective weight maps. Astrometry corrections are done by the pipeline by aligning stars from GAIA DR2 and the 2MASS catalog. The HELMS-55 environment only has coverage in the F110W band, and the SDP.17b environment has coverage in the F110W and F160W bands, with identical footprints. The SDP.17b and HELMS-55 HST data were originally presented in \cite{Negrello_2014} and \cite{Borsato_2024}, respectively. The footprints of the ancillary data are shown in Figure \ref{fig:SDP17_HELMS-55_data}.

\section{Methods}\label{sec:Methods}

\subsection{Source Detection \& Photometry}\label{subsec:detection_and_photometry}

Source detection and photometry were performed using the software \href{https://www.astromatic.net/software/sextractor/}{\texttt{SExtractor}} \citep{SExtractor} in dual image mode. The main \texttt{SExtractor} parameters used are summarized in Appendix \ref{subsec:source_ext_and_phot}, following standard choices adopted in the literature (e.g., \citealt{Postman_2012, Estrada_2023}). The $K_s$-band images were used as the detection images, while measurements were performed on images of all available filters as listed in Table \ref{tab:photometric_data}. The ancillary images were resampled to match the pixel scale of the $K_s$-band image. To account for the noise variation across pixels, the uncertainty maps produced by \texttt{DRAGONS} were used for the $K_s$-band images, while weight maps were retrieved, as described in Section \ref{subsec:ancillary_data}, for the other bands used.

Regions in the $K_s$-band images were masked out for source detection and photometry as follows. Both target environments' $K_s$-band images contain imaging artifacts due to the F2 On-Instrument Wavefront Sensor (OIWFS) probe. These artifacts were identified in the calibrated images as regions with saturated pixel values ($\gtrsim 10^4$ ADU). Additionally, the regions outside the F2 circular imaging FOV contained only low-ADU pixels and were free of astronomical sources. These regions contained unusable data and were masked. About 30\% of the spatial extent, including regions outside the F2 circular imaging FOV, of the $K_s$-band images were masked. The usable area after masking is shown in Table \ref{tab:photometric_data} and Figure \ref{fig:SDP17_HELMS-55_data}.

The fluxes of sources were measured within circular apertures measuring 2.5$''$ in diameter. This aperture was chosen since it maximized the number of detections in both target environments (the largest aperture before noise-dominated flux measurements). A detection was defined as having a signal-to-noise ratio (SNR) $\geq 3$ in aperture flux. We discard sources with SNR $<3$ in the $K_s$-band, corresponding to 8 and 5 sources in SDP.17b and HELMS-55 environments, respectively. Sources with aperture photometry in the $K_s$-band biased by neighboring sources were discarded based on the flags from \texttt{SExtractor}. This resulted in 42 and 32 additional sources being discarded from the SDP.17b and HELMS-55 target environments, respectively. The total number of sources detected in each target environment after discarding sources based on the $K_s$-band SNR and the \texttt{SExtractor} flags are listed in Table \ref{tab:realizations_compared_to_COSMOS}. We detect 1,192 sources based on these criteria (1,114 excluding known stars and the central DSFGs).

\begin{table*}[ht]
    \centering
    \begin{tabular}{c|ccc|cc|cc|c}
        \toprule
        \textbf{Target} & \textbf{Sources} & \textbf{Known} & \textbf{ML Companion} & $\boldsymbol{\overline{N}_\text{target}}$ & $\boldsymbol{\overline{N}_\text{COSMOS}}$ & $\boldsymbol{S}$ & $\boldsymbol{\delta}$ & $\boldsymbol{M_\text{h, lim}(z)}$ \\
        \textbf{Environment}  & \textbf{Detected} & \textbf{Stars} & \textbf{Galaxies} & & & \textbf{[$\sigma_\text{COSMOS}$]} & & \textbf{[$10^{13} M_\odot$]} \\
        \midrule
        SDP.17b   & 526 & 49 & 2  & $5 \pm 2$& $4 \pm 3$ & $0.2 \pm 0.4$& $0.1 \pm 0.2$ (59\%)& $\geq 0.06 \pm 0.01$\\
        HELMS-55 & 666 & 27 & 13 & $15 \pm 3$ & $7 \pm 3$ & $2.2 \pm 0.6$ & $1.0 \pm 0.3$ (99\%) & $\geq 1.2 \pm 0.1$\\
        \bottomrule
    \end{tabular}
    \caption{The total number of sources detected (including stars), known stars identified, maximum likelihood (ML) companion galaxies, and average number of companion galaxies ($\overline{N}_\text{target}$) identified in each target environment. Sources flagged by \texttt{SExtractor} as having $K_s$-band aperture photometry biased by neighboring sources were discarded, as were sources with $K_s$-band SNR $<3$. The number of companion galaxies is shown after applying the $K_s$-band $3\sigma$ point-source sensitivity cut within a $2''$ diameter aperture. $\overline{N}_\text{COSMOS}$ represents the average number of field galaxies expected for the F2 FOV, estimated from the COSMOS survey. The overall significance ($S$) of $\overline{N}_\text{target}$ compared to $\overline{N}_\text{COSMOS}$ is given in units of $\sigma_\text{COSMOS}$ (the standard deviation of $N_\text{COSMOS}$). The galaxy overdensity ($\delta$) is reported for each target environment, with the percentage of realizations yielding $\delta > 0$ shown in brackets. $M_\text{h, lim}(z)$ is the lower limit of the halo mass of the overdensities at the spectroscopic redshift of the central DSFG.}
    \label{tab:realizations_compared_to_COSMOS}
\end{table*}

Stars were used to perform flux calibration and determine the PSF FWHMs of each band (see Table \ref{tab:photometric_data}). $K_s$-band detections were cross-matched with the GAIA DR3 (\hspace{-0.06cm}\citealt{GAIA_2023}) and 2MASS \citep{2MASS_2006} catalogs to flag known stars. The number of known stars identified in each target environment is listed in Table \ref{tab:realizations_compared_to_COSMOS}. The FWHMs for each band were calculated by fitting a 2D Gaussian to a stacked cutout of known stars within the respective image. Aperture corrections for the $K_s$ and HSC bands were derived from their respective curves of growth. The HST bands' aperture corrections were determined using the curves of growth provided by the STScI HST documentation\footnote{\url{https://www.stsci.edu/hst/instrumentation/wfc3/data-analysis/photometric-calibration/ir-encircled-energy}}. 
For each band, the 3$\sigma$ point-source sensitivity was estimated by first computing the standard deviation of fluxes measured in 1000 randomly placed $2''$ diameter circular apertures in source-free regions. This standard deviation was multiplied by three to obtain the 3$\sigma$ flux limit, which was then converted to an AB magnitude. The $2''$ aperture was chosen to compare to the $2''$ diameter aperture flux measurements of the COSMOS survey (\citealt{COSMOS_2022}; see Section \ref{subsubsec:field_galaxies} for further details).

The $K_s$-band zeropoint for each target environment was estimated using the 2MASS $K_s$-band 4$''$ (in radius) aperture magnitudes of the known stars.  Aperture corrections were applied to the $K_s$-band measurements to match the 2MASS aperture size before deriving the zeropoint. The zeropoints for the \textit{grizy} bands of each target environment were estimated using the HSC PDR3 5.7$''$ (in diameter) aperture magnitudes of the known stars. HST zeropoints were obtained via the header of the respective images. Table \ref{tab:photometric_data} contains the zeropoint values.

\subsection{Photometric Redshifts}\label{subsec:phot_z}

We estimated photometric redshifts, $z_\text{phot}$, for the 1,192 sources detected in the SDP.17b and HELMS-55 environments using the \href{https://github.com/gbrammer/eazy-py}{\texttt{EAZY-PY}} software package \citep{Brammer_2008}, applied to the photometric catalogs generated as described in Section \ref{subsec:detection_and_photometry}. \texttt{EAZY-PY} fits a non-negative linear combination of templates onto observed spectral energy distributions (SEDs) via a least-squares fit. The default 12 flexible stellar population synthesis (FSPS; \citealt{Conroy_Gunn_White_2009, Conroy_Gunn_2010}) templates were chosen to span a wide range of galaxy types (e.g., star-forming, quiescent, dusty). These templates assume a \citet{Chabrier_2003} IMF, \cite{Kriek_Conroy_2013} dust attenuation law, and solar metallicity. A systematic flux error of 10\% was assumed to reflect the calibration uncertainties of the ground-based imaging data. As done in \citet{Weaver_2024}, we do not apply a UV slope prior, apparent magnitude prior, or iterative corrections to the zeropoints. Each source has a $z_\text{phot}$ probability density function (PDF), $p(z)$, outputted. Following \cite{Stevans_2021}, sources with reduced chi-squared values $\chi^2_r \geq  10$, were discarded from further analysis.

To judge the accuracy of the photometric redshift estimates, we compared the $z_\text{phot}$ to $z_\text{spec}$ for sources in the SDP.17b and HELMS-55 environments. HSC PDR3 provides $z_\text{spec}$ measurements for four sources in total, with three in SDP.17b's environment and one in HELMS-55's environment. In addition, a query of the Sloan Digital Sky Survey (SDSS) DR18\footnote{\url{https://skyserver.sdss.org/dr18/SearchTools/sql}} \citep{SDSS_DR18_2023} yielded eight more sources with spectroscopic redshifts in HELMS-55's environment. This gave a combined sample of twelve sources with $z_\text{spec}$. From this set, we excluded two SDSS stars with $z_\text{spec} < 0.01$, as these values fall below the minimum $z_\text{phot}$ value (0.01) in the \texttt{EAZY-PY} redshift grid. We also removed two quasi-stellar objects (QSOs; indicated on the SDSS spectra results) because their photometry is dominated by AGN emission, preventing accurate $z_\text{phot}$ estimates. The remaining eight sources have photometric redshifts within $1\sigma$ of their $z_\text{spec}$ in the two target environments, with a median $1\sigma$ error of $\sim 0.2$. All these $z_\text{spec}$ are low redshift ($z \lesssim 1$) sources and, therefore, are not companions to the DSFGs. This subset of sources with $z_\text{spec}$ measurements included known stars. Hereafter, known stars were excluded from further analysis, and we considered the rest of the sources as candidate galaxies.

\begin{table*}[ht]
\centering
\begin{tabular}{ccccccccc}
\toprule
\textbf{ID}&  \textbf{RA}& \textbf{Dec} &$\boldsymbol{z_\text{ml}}$&$\boldsymbol{z_\text{16}}$ &$\boldsymbol{z_\text{50}}$ &$\boldsymbol{z_\text{84}}$ &$\boldsymbol{K_s}$\textbf{-band}  &\textbf{Distance to DSFG}\\ 
&  [degrees]& [degrees] && & & &\textbf{Mag}&[cMpc]\\ 
\midrule
\multicolumn{9}{c}{\textbf{SDP.17b Environment}}\\
\midrule
174& 135.7342777& -1.7064637& 2.34& 1.48& 2.24& 3.08& 22.80 $\pm$ 0.08&3.23\\
604& 135.7511012& -1.7010284& 2.14& 1.89& 2.18& 2.45& 22.54 $\pm$ 0.06&1.53\\
\midrule
\multicolumn{9}{c}{\textbf{HELMS-55 Environment}}\\
\midrule
51&  352.0924886&  -0.6515021&2.24 & 1.67& 3.05&4.61 &22.46 $\pm$ 0.05 &4.69\\ 
103&  352.1200005&  -0.6385473&2.24 & 1.51& 2.36&2.97 &22.01 $\pm$ 0.03 &3.96\\ 
175&  352.1402265&  -0.6330804&2.25 & 1.52& 3.07&4.72 &22.45 $\pm$ 0.05 &4.38\\ 
198&  352.1311948&  -0.6482481&2.32 & 1.37& 2.30&4.01 &21.48 $\pm$ 0.02 &2.80\\ 
210& 352.1135123& -0.6674607&2.25 & 1.08& 2.31&3.52 &22.97 $\pm$ 0.08 &2.09\\ 
236& 352.1543876&   -0.638105&2.23 & 1.38& 2.17&2.58 &21.00 $\pm$ 0.01 &4.39\\ 
247& 352.1513104&  -0.6362687&2.28 & 1.37& 2.25&2.80 &22.87 $\pm$ 0.07 &4.41\\ 
273& 352.1056724&  -0.6888228&2.48 & 1.73& 2.39&2.86 &21.43 $\pm$ 0.02 &2.94\\ 
279& 352.1256695&  -0.6686597&2.21 & 1.36& 3.21&5.05 &23.1 $\pm$ 0.1 &1.03\\ 
314&  352.139979&  -0.6608564&2.11 & 1.32& 1.77&2.14 &21.04 $\pm$ 0.01 &1.71\\
351& 352.0968222&  -0.7077949&2.30 & 1.78& 2.97&4.24 &22.47 $\pm$ 0.05 &4.73\\
499& 352.1474863&  -0.7028016&2.25 & 1.79& 2.50&3.36 &21.70 $\pm$ 0.03 &3.02\\
525& 352.1343332&  -0.7229986&2.42 & 1.88& 2.52&3.23 &22.09 $\pm$ 0.04 &4.64\\
591& 352.1307794&  -0.7115472&2.42 & 1.32& 2.08&2.51 &21.09 $\pm$ 0.01 &3.50\\
\bottomrule
\end{tabular}
\caption{The coordinates and maximum likelihood photometric redshifts ($z_\text{ml}$) of SDP.17b and HELMS-55's maximum likelihood companion galaxies, before applying a magnitude cut based on the target environment's $K_s$-band $3\sigma$ point-source sensitivities within a $2''$ diameter aperture (see Table \ref{tab:photometric_data}). Additionally, the 16$^{th}$, 50$^{th}$, and 84$^{th}$ percentile $z_\text{phot}$ are also given. All companion galaxies are within $dz=0.2$ of their DSFG's $z_\text{spec}$. The $K_s$-band magnitude measured within a $2''$ diameter aperture is also provided. The final column lists the projected distance to the respective DSFG.} 
\label{tab:comapnions}
\end{table*}

\subsection{Overdensity Analysis}\label{subsec:overdensities}

To measure the overdensity in the target environments, we adopted a statistical approach to estimate both the number of companion galaxies around the DSFGs and the average number of field galaxies expected within the F2 FOV. The number of companion galaxies was estimated through Monte Carlo sampling the $p(z)$. This robustly samples the posterior distribution of each $z_\text{phot}$ fit, providing a statistically representative estimate of the true galaxy population within the target environments. A control sample of field galaxies was generated from the COSMOS survey \citep{COSMOS_2022} by sampling random apertures the size of the F2 FOV, providing an estimate of the expected field population. This minimized cosmic variance by averaging over multiple independent sightlines in a well-studied, large-area survey, thereby reducing biases from localized density fluctuations. A magnitude cut equal to the $K_s$-band $3\sigma$ point-source sensitivity (within a $2''$ diameter aperture) of each target environment was applied to both the companion galaxies and the COSMOS control sample of field galaxies. Matching the depth of both datasets ensured a fair comparison between the target environments and the control sample.

We summarize the main steps of the overdensity analysis below before presenting the detailed methods in the following subsections.
\begin{enumerate}
\item \textbf{Sample redshifts:} For each source in the F2 FOV (excluding the central DSFG), 1000 Monte Carlo realizations are performed, each drawing a random redshift ($z_\text{r}$) from its photometric redshift probability distribution, $p(z)$.
\item \textbf{Identify companion galaxies:} Classify a source as a companion galaxy if $|z_\text{r} - z_\text{DSFG}| \leq 0.2$ and it is brighter than or equal to the $K_s$-band $3\sigma$ point-source sensitivity (within a $2''$ diameter aperture) of the target environment. Record the number of companion galaxies per realization (${N_\text{target}}$), then compute the mean and standard deviation across all realizations, as described in Section \ref{subsubsec:companion_galaxies}.
\item \textbf{Build control sample:} From the COSMOS survey, select galaxies meeting the same redshift and magnitude criteria, masking regions with unreliable photometry (e.g., near bright stars, image edges, or lacking NIR coverage). Draw 1000 random apertures with the same usable area as the $K_s$-band F2 FOV. Compute the mean ($\overline{N}_\text{COSMOS}$) and standard deviation of galaxy counts ($\sigma_\text{COSMOS}$) in the random apertures (see Section \ref{subsubsec:field_galaxies}).
\item \textbf{Evaluate the overdensity:} Compare ${N}_\text{target}$ to $\overline{N}_\text{COSMOS}$ to determine both the significance and contrast of the overdensity, as detailed in Section \ref{subsubsec:Emprical_overdensities}.
\end{enumerate}

\subsubsection{Companion Galaxies}\label{subsubsec:companion_galaxies}

To perform a statistical analysis of the surface density of photometric sources, we sampled the posterior distribution of each source's $p(z)$ via Monte Carlo sampling. In each realization, sources were assigned redshifts, $z_r$, randomly drawn from their $p(z)$ produced by \texttt{EAZY-PY}. For each target environment, 1000 realizations were generated, and galaxies were classified as companion galaxies if $|z_\text{r} - z_\text{DSFG}| \leq 0.2$, where $z_\text{DSFG}$ is the spectroscopic redshift of the central DSFG of the environment. Three $dz$ options were tested: $0.1, 0.2, 0.3$. A fiducial $dz=0.2$ was adopted, as it provided a good balance between including likely companions and minimizing interlopers. 
We follow a similar approach to that of \cite{Kiyota_2024}, adopting $dz = 0.2$, to identify overdensities. For comparison, \cite{Castignani_2014} used redshift bins of $\Delta z  =0.02 - 0.4$ (i.e. $dz = \Delta z/2 = 0.01 - 0.2$) to search for high-redshift protoclusters at $z=0.4-4.0$ using photometric redshifts. Similarly, \cite{Calvi_2023} used an average redshift bin of $\overline{\Delta} z =0.3$ (i.e., $dz = \overline{\Delta} z/2 = 0.15$) while searching for protoclusters around DSFGs using $z_\text{phot}$. 

\begin{figure*}
    \centering
    \includegraphics[width=\linewidth]{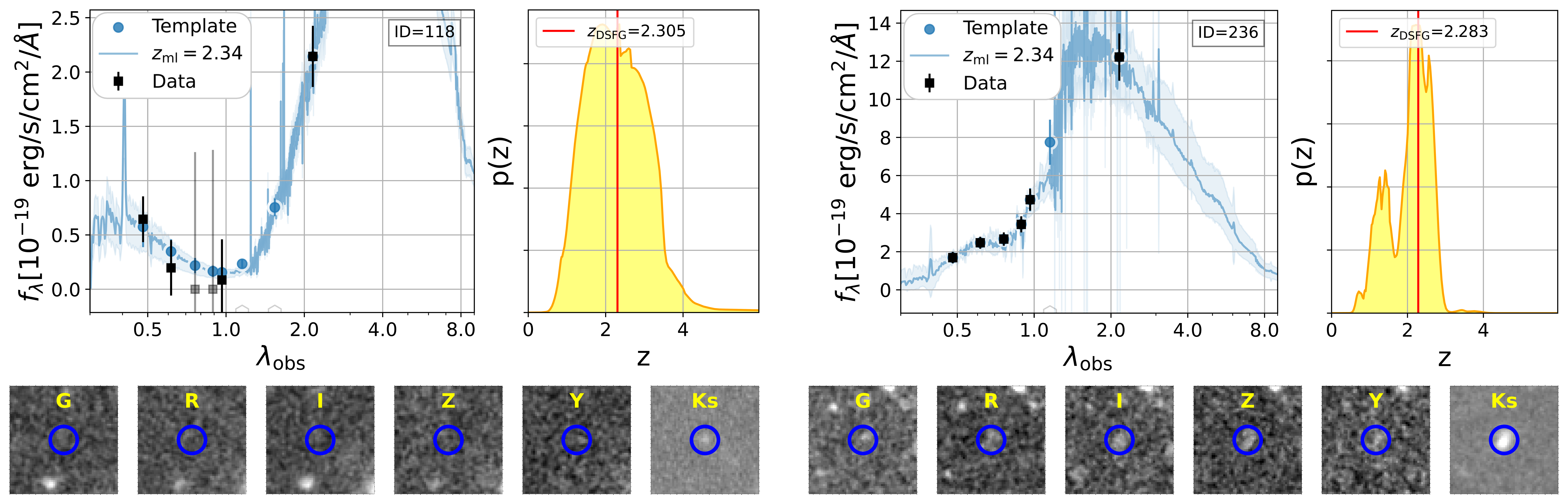}
    \caption{Example \texttt{EAZY-PY} SED fits of maximum likelihood  companion galaxies. The left and right panels correspond to a companion galaxy from the SDP.17b and HELMS-55 target environments, respectively. Each panel shows the SED plot on the left and the probability distribution function of the photometric redshift, $p(z)$, with respect to the redshift on the right. The $z_\text{spec}$ value of the central DSFG is shown in red in the $p(z)$ plots, and the maximum likelihood $z_\text{phot}$ estimate, $z_\text{ml}$, is used to compute the best-fit SEDs, which are plotted in blue, with the $1\sigma$ error shaded around them. Template fluxes are shown as blue circles, while observed fluxes and their error bars are shown in black (SNR $> 0$) and gray (SNR $\leq 0$). Open markers at the bottom of the SED panels mark bands without flux measurements (e.g., HST F110W and F160W). For each source, cutouts measuring $10''$ by $10''$ are shown in the available filters, with $2.5''$ in diameter apertures overlaid, and filter names shown in yellow (e.g., \textit{grizy}, $K_s$).}
    \label{fig:example_SEDs_zml}
\end{figure*}

The maximum likelihood companion galaxies were identified from the $p(z)$ produced by the \texttt{EAZY-PY} SED fits, ensuring their redshifts lie within $dz = 0.2$ around the central source  ($|z_\text{ml} - z_\text{DSFG}| \leq 0.2$, where $z_\text{ml}$ is the maximum likelihood redshift; see Figure \ref{fig:example_SEDs_zml} for two example sources). These galaxies were used to estimate SFR and $M_\star$ values representative of each target environment (see Section \ref{subsec:SFR_M_cigale}). The coordinates, $z_\text{phot}$, $K_s$-band magnitudes, and projected distance from the central DSFG for the maximum likelihood companion galaxies are listed in Table \ref{tab:comapnions}. \texttt{EAZY-PY} SED fits and cutouts of these galaxies are shown in Appendix \ref{subsubsec:companion_galaxies_before_mag_cut}.

Each Monte Carlo realization yielded a $N_\text{target}$ value, representing the number of companion galaxies in the target environment. For each target environment, the average $\overline{N}_\text{target}$ and its standard deviation were recorded as the overall number of companion galaxies and the corresponding uncertainty, respectively.  The number of companion galaxies identified in the SDP.17b and HELMS-55 environments is $5 \pm 2$ and $15 \pm 3$, respectively, when applying the magnitude cut (Table \ref{tab:realizations_compared_to_COSMOS}), and $6 \pm 2$ and $16 \pm 4$, respectively, when no magnitude cut is applied (Table \ref{tab:realizations_no_mag_cut_numbers}).

\subsubsection{Control Sample of Field Galaxies}\label{subsubsec:field_galaxies}

The COSMOS survey \citep{COSMOS_2022}, which covers a wide area of $\sim 2 \, \text{deg}^2$, was used as a proxy for the average number of expected field galaxies. Only COSMOS sources flagged as galaxies in the \texttt{Classic} COSMOS catalog \citep{COSMOS_2022} were included. Galaxies located near bright stars, at the edges of HSC/Suprime-Cam images, or in regions without UltraVISTA NIR coverage were masked using \texttt{FLAG\_COMBINED}=0. Sources with neighbors close and bright enough to bias the photometry were also masked using the provided \texttt{SExtractor} flags. To ensure a consistent galaxy selection in the COSMOS survey and the target environments, a magnitude cut in the $K_s$-band was applied. Only sources brighter than or equal to the $K_s$-band sensitivities in Table \ref{tab:photometric_data} were included in the overdensity analysis. Additionally, we apply the same redshift cut as the companion galaxies, $| z_\text{ml} - z_\text{DSFG} | \leq 0.2$ where $z_\text{ml}$ is the \texttt{EAZY-PY} maximum likelihood photometric redshift, as provided in the COSMOS catalog, and discard sources with \texttt{EAZY-PY} $\chi_r \geq 10$. We refer to this subsample of galaxies as the COSMOS control sample hereafter. 

The average number of field galaxies expected within a F2 FOV, $\overline{N}_\text{COSMOS}$, was estimated by taking 1000 random circular apertures in the COSMOS control sample and averaging the number of galaxies found within them. Each aperture had the same area as the $K_s$-band usable area reported in Table \ref{tab:photometric_data}. For $\overline{N}_\text{COSMOS}$, apertures were not placed in areas masked by \texttt{FLAG\_COMBINED}. As done in Section \ref{subsubsec:companion_galaxies}, the standard deviation of $\overline{N}_\text{COSMOS}$ was reported as its uncertainty. The $\overline{N}_\text{COSMOS}$ values and their uncertainties are reported in Table \ref{tab:realizations_compared_to_COSMOS}.

It is worth noting that the COSMOS survey contains several protoclusters at $z\sim 2.3$ (see Table 1 of \citealt{Ata_2022} for a summary) as well as several proto-groups (see \citealt{Toni_2025}). These structures could potentially bias the average number of field galaxies found in the COSMOS survey, leading to an underestimated overdensity in the SDP.17b and HELMS-55 environments. However, fewer than $0.5\%$ of the random circular apertures used in our analysis fall within 3.1$'$ of the known protoclusters (based on the F2 FOV radius) reported in \citet{Ata_2022}, suggesting the contamination is minimal.

\subsubsection{Empirical Overdensity and Significance}\label{subsubsec:Emprical_overdensities}

We compute two commonly used measures of galaxy overdensity to facilitate comparison with the literature. As a first measure of overdensity, we compute the statistical significance of the number of companion galaxies as follows,
\begin{equation}
S_i = \frac{{N}_{\text{target}, i} - \overline{N}_\text{COSMOS}}{\sigma_\text{COSMOS}},
\label{eq:empirical_signifigance}
\end{equation}
where $S_i$ is the overdensity significance and ${N}_{\text{target}, i}$ is the number of companion galaxies identified for realization $i$ from the Monte Carlo sampling. $\overline{N}_\text{COSMOS}$ and $\sigma_\text{COSMOS}$ are the average and standard deviation of the expected number of galaxies within the COSMOS control sample, respectively. 

In addition, the galaxy overdensities in the target environments were computed by comparing the number of companion galaxies identified from the Monte Carlo sampling to the average number of field galaxies from the COSMOS control sample as follows,
\begin{equation}
\delta_i = \frac{{N}_{\text{target}, i}} {\overline{N}_\text{COSMOS}} -1,
\label{eq:empirical_overdensity}
\end{equation} 
where $\delta_i$ is the empirical overdensity estimate for realization $i$ of the target environment. The percentage of realizations being overdense with $\delta_i > 0$ is reported in Table \ref{tab:realizations_compared_to_COSMOS}.

Both the overdensity $\delta$ and the significance $S$ capture complementary information. The former describes the relative strength of the excess of galaxy counts, indicating how much denser the target environments are compared to the field galaxy population. In contrast, $S$ reflects the statistical confidence that the excess is real, rather than a product of random fluctuations. To avoid a bias in the overdensity analysis, the central DSFGs were excluded from the overdensity $S$ and $\delta$ calculations. We report the overall $S$ and $\delta$ values for each target environment in Table \ref{tab:realizations_compared_to_COSMOS}, using the median and median absolute deviation (MAD) of the $S_i$ and $\delta_i$ distributions to account for the moderate non-Gaussianity observed in the companion galaxy distributions from the Monte Carlo sampling of the SDP.17b and HELMS-55 target environments ($\sigma_\text{target}$/MAD = 2.11 and $\sigma_\text{target}$/MAD = 1.73, respectively).

\subsection{Galaxy Properties: CIGALE SED Fitting}\label{subsec:SFR_M_cigale}

Using the SED fitting software \texttt{CIGALE} \citep{Boquien_2019}, we estimate the SFRs and stellar masses of the maximum likelihood companion galaxies in the SDP.17b and HELMS-55 environments. No magnitude cut was applied, as we are not directly comparing number counts to the COSMOS survey. Incompleteness is not a major concern in this context, since the main purpose is to assess relative positions of companion galaxies with respect to the star-forming main sequence. Although SFR and $M_\star$ estimates were provided in the \texttt{EAZY-PY} fits described in Section \ref{subsec:phot_z}, they were not used in our analysis because \texttt{EAZY-PY} FSPS templates assume a simple star formation history (SFH). 

\texttt{CIGALE} uses a Bayesian approach to estimate galaxy properties by fitting SED models to observed photometry. The spectroscopic redshifts of SDP.17b and HELMS-55 were used as fixed redshifts for their respective maximum likelihood companion galaxies. However, \texttt{CIGALE} does not account for uncertainties in the input redshift, and as a result, the uncertainties on derived quantities such as SFR and $M_\star$ are likely underestimated \citep{Acquaviva_2015}. Given that our sources are selected with photometric redshifts within $dz = 0.2$, we estimate that this redshift uncertainty contributes approximately 0.09 dex to the uncertainty in SFR and $M_\star$, based on their dependence on luminosity distance at $z \sim 2.3$. To conservatively account for this, we add a minimum uncertainty of 0.1 dex in quadrature to the SFR and $M_\star$ uncertainties. SFR and stellar mass estimates were not performed on the DSFGs SDP.17b and HELMS-55 due to the foreground lens contamination in their photometry. As done in Section \ref{subsec:phot_z} for the $z_\text{phot}$ estimates, a systematic flux error of 10\% was added to reflect the calibration uncertainties of the ground-based imaging data. These SED models are generated by specifying prior distributions over a range of parameter values, which are then used in various modules (e.g., SFH, stellar populations, and dust attenuation). 
For SED modeling, we adopted a delayed SFH,
\begin{ceqn}
\begin{equation} 
\text{SFR}(t) \propto  \frac{t}{\tau^2} e^{-t/\tau}, \, \text{for} \, 0 \leq t \leq t_0,
\end{equation}
\label{eq:sfr}
\end{ceqn}
where $\text{SFR}(t)$ is given in units of $\text{M}_\odot  \,\text{yr}^{-1}$, $t$ is the time since the onset of star formation, $t_0$ is the lookback time of the onset of star formation, and $\tau$ is the e-folding timescale of star formation. A delayed SFR can model both early-type (small $\tau$) and late-type galaxies (large $\tau$; \citealt{Boquien_2019, Tacchella_2022, Donevski_2023, Ronconi_2024}). We assumed a \citet{Chabrier_2003} IMF, fixed solar metallicity, and the \texttt{dustatt\_modified\_starburst} dust attenuation law based on \citet{Calzetti_2000}, extended to far-UV using \citet{Leitherer_2002}. The Calzetti law was chosen since \cite{Hamed_Darko_2023} demonstrated that the curve provides a good fit for galaxies of diverse stellar masses around the cosmic noon. Stellar emission was modeled with the \citet[BC03]{Bruzual_Charlot_2003} library. The specific parameters and modules used to generate the SED models are listed in Table \ref{tab:cigale_param}, and a full description of each module can be found in \citet{Boquien_2019}.

\subsection{Halo Mass Estimates at the Redshifts of the DSFGs}\label{subsec:halo_mass}

Another key aspect of the DSFG environments is the dark matter halo mass associated with the overdensities. This helps determine if the overdensities are massive enough to be considered protoclusters. Following \citet{Laporte_2022} and \citet{Brinch_2023}, we estimate the halo mass by converting the total stellar mass of the maximum likelihood companion galaxies to halo mass with the baryonic-to-dark matter fraction from \citet{Planck_2018_2020_paper},
\begin{ceqn}
\begin{equation} 
M_\text{h, lim}(z) = M_{\star, \text{tot}}(z) \left[\Omega_\text{c} / \Omega_\text{b} \right]
\label{eq:stelar_to_halo_mass}
\end{equation}
\end{ceqn}
where $ M_{\star, \text{tot}}$ is the sum of the stellar masses of the maximum likelihood companion galaxies, $\Omega_\text{c} h^2 = 0.120 \pm 0.001$ and $\Omega_bh^2=0.0224 \pm 0.0001$ are the cold dark matter and baryon densities, respectively. This method assumes all companion galaxies are identified and does not account for gas or the intracluster medium, therefore, the halo mass is a lower limit.  We estimate uncertainties by propagating errors in $M_\star$ from \texttt{CIGALE} and the density parameters from \citet{Planck_2018_2020_paper}. The halo mass estimates and uncertainties are reported in Table \ref{tab:realizations_compared_to_COSMOS}, with further comparison to known protoclusters and simulations in Section \ref{sec:results_and_implications}.

\section{Results \& Discussions}\label{sec:results_and_implications}

\subsection{Are DSFGs Protocluster Signposts?}\label{subsec:dsfgs_lighthouses}

DSFGs have long been proposed as promising signposts for identifying protoclusters \citep{Capak_2011, Umehata2015, Miller_2018, Oteo_2018, Harikane_2019, Long_2020, Calvi_2023, Araya_Araya_2024}, due to their tendency to reside in massive dark matter halos \citep{Marrone_2018, Garcia_Vergara_2020, Stach_2021, Araya_Araya_2024}. However, whether DSFGs consistently trace the most overdense regions of the early Universe remains uncertain \citep{Chapman_2009, Chapman_2015, Miller_2015, Casey_2016, Alvarez_2021, Gao_2022}.

\begin{figure*}
    \centering
    \includegraphics[width=0.85\linewidth]{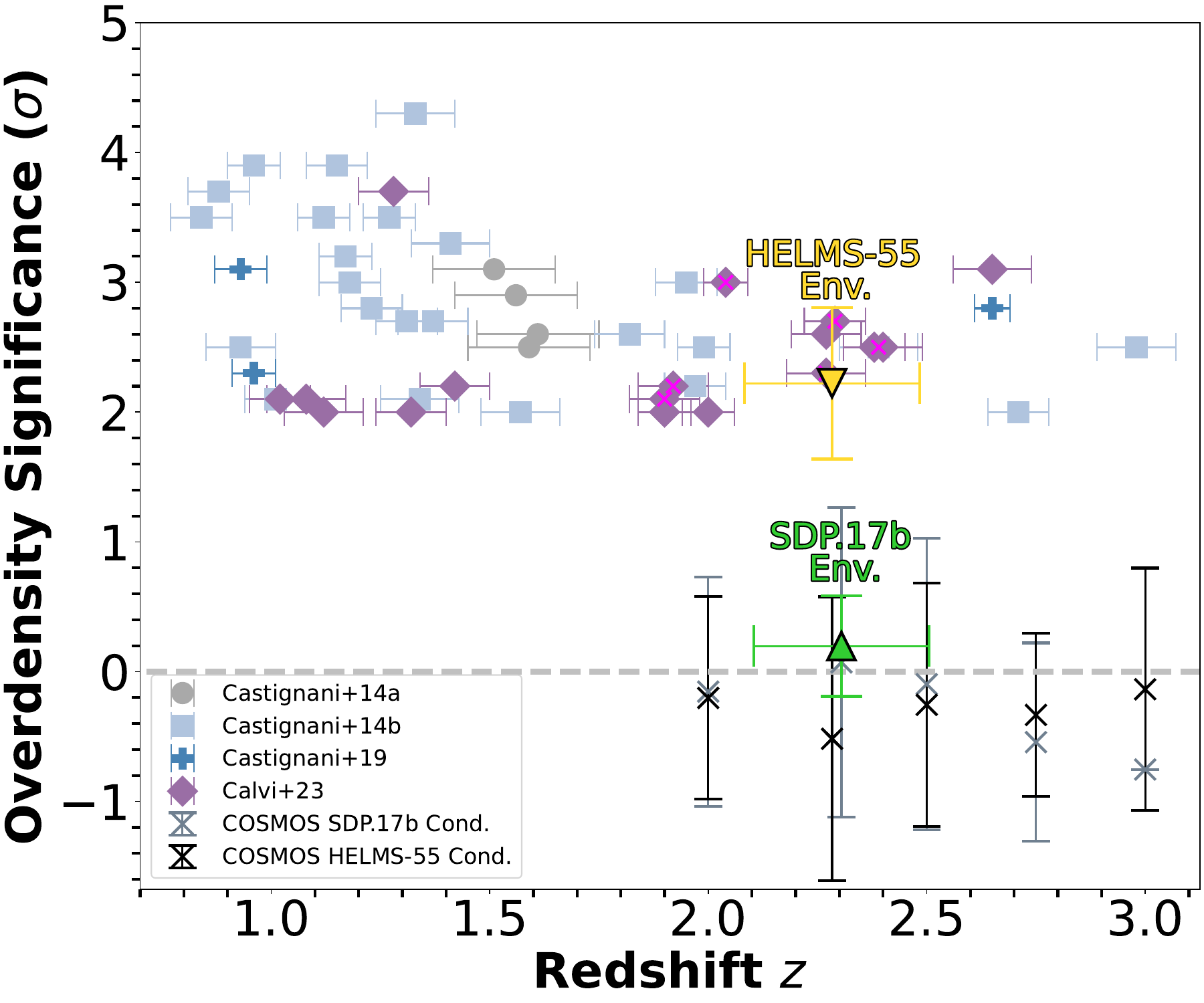}
    \caption{Overdensity significances as a function of redshift for galaxy cluster and protocluster candidates photometrically identified in the literature. The overdensity significances of the SDP.17b and HELMS-55 environments are plotted in comparison as green and yellow triangles, respectively. Symbols represent different studies: gray circles for \citet{Castignani_2014}, light-blue squares for \citet{Castignani_2014_radio}, blue pluses for \citet{Castignani_2019}, and purple diamonds for \citet{Calvi_2023}. However, these papers do not report error bars on their overdensity significance. Spectroscopically confirmed protoclusters are marked with magenta crosses. The median value of the overdensity significance, assuming the number of companion galaxies in a given realization is equal to the number of field galaxies drawn from the COSMOS control sample ($N_{\text{target}, i} = N_{\text{COSMOS}, i}$), is shown as gray and black crosses at the $K_s$-band magnitude cuts of the SDP.17b and HELMS-55 environments, respectively.}
    \label{fig:lit_comparrison}
\end{figure*}

\subsubsection{What Are the Environmental Densities Around Our DSFGs?}

We find that our two DSFGs at $z \sim 2.3$ reside in contrasting environments. We identified more companion galaxies than expected within the target environments compared to the COSMOS control sample. The target environments show both low and moderate overdensities, with values ranging from $\delta = 0.1$ to $1.0$, corresponding to significance levels between $0.2\sigma$ and $2.2\sigma$, as listed in Table \ref{tab:realizations_compared_to_COSMOS} where $\sigma$ refers to the standard deviation of the COSMOS control sample. One DSFG, HELMS-55, resides in an overdensity consistent with a protocluster candidate, with 15 $\pm$ 3 companion galaxies, ranging from 1.03 to 4.73 cMpc in projected distance from the DSFG. The other DSFG, SDP.17b, is surrounded by only 5 $\pm$ 2 companion galaxies, ranging from 1.53 to 3.23 cMpc in projected distance from the DSFG, consistent with a typical field environment. These projected distances are based on the maximum likelihood companion galaxies identified, listed in Table \ref{tab:comapnions}. The number of companion galaxies prior to the magnitude cut made for our overdensity analysis is provided in Table \ref{tab:realizations_no_mag_cut_numbers}. These findings suggest that while some DSFGs can trace protocluster regions, they may not universally mark the most overdense environments.

Only the overdensity significance of the HELMS-55 environment is consistent with known protocluster candidates and spectroscopically confirmed protoclusters in the literature at $z \sim 2.3$, as shown in Figure \ref{fig:lit_comparrison}. Prior studies have shown that galaxy overdensities at or above $2.0\sigma$ are often associated with real, large-scale structures (e.g., \citealt{Chapman_2009, Galametz_2013, Diener_2015, Oteo_2018}). These include systems traced by DSFGs and radio galaxies across $1 \lesssim  z \lesssim 5$. The typical galaxy overdensity for protoclusters at $z \sim 2$ varies in the literature, often depending on the method used for estimation (see \citealt{Muldrew_2012} for a detailed comparison of different methods). For example, the PKS 1138-262 protocluster candidate at $z=2.16$ has a galaxy overdensity of $\delta = 3 \pm 2$ \citep{Venemans_2007, Chiang_2013}, while J2143-4423 at $z=2.38$ has an average overdensity of $\delta = 4.8 \pm 2.5$ \citep{Palunas_2004}. Both overdensities were estimated by comparing the number density of Ly$\alpha$ emitters in blank fields from studies such as \citet{Hayashino_2004} and \citet[see Table 5 of \citealt{Chiang_2013} and references therein]{Cowie_1998AJ....115.1319C}. Given the variation in estimation methods, we compare the significance ($S$) of our overdensities to those reported for known protoclusters, rather than directly comparing $\delta$ values. We reach qualitatively similar conclusions when comparing our $\delta$ values for both target environments to those of protocluster candidates such as PKS 1138-262 and J2143-4423.

Both SDP.17b and HELMS-55 lie at sufficiently high redshift ($z > 1.5$), where observations and simulations suggest that most halos are still in the process of collapsing and have not yet virialized into massive clusters \citep{Chiang_2013, Muldrew_2015, Overzier2016}. HELMS-55 has a high probability of being overdense, with $99$\% of Monte Carlo realizations returning an overdensity relative to the COSMOS survey. In contrast, SDP.17b is classified as overdense in only 59\% of the realizations.

Moreover, based on our halo mass estimates in Table \ref{tab:realizations_compared_to_COSMOS}, and the halo mass–redshift relationship from \citet[Figure 18]{Behroozi_2013}, we estimate that the SDP.17b and HELMS-55 environments will evolve into halos with masses of at least $\sim10^{13} M_\odot$ and $\sim10^{14} M_\odot$, respectively, by $z=0$, with an estimated scatter of 0.25 dex. Only the HELMS-55 environment halo mass is consistent with the $\sim10^{14}\,M_\odot$ lower bound typically quoted for the halo masses of galaxy clusters at $z=0$. Additionally, the target environments have halo masses consistent with those expected for submillimeter-selected galaxies, based on clustering observations in SCUBA-2 surveys. The surveys find that DSFGs at $z \sim 2-3$ typically reside in dark matter halos of $M_\text{halo} \sim 10^{12.8} - 10^{13}\, M_\odot$ \citep{Wilkinson_2017, Stach_2021}. Based on Millennium-based cluster evolution models \citep{Chiang_2013}, the HELMS-55 environment halo mass falls within the range expected for Fornax-type clusters.

All together, the overdensity significances, probability of the target environments being overdense, and halo mass estimates suggest HELMS-55 resides in a protocluster candidate, while the SDP.17b environment appears consistent with a typical field environment. The contrasting environments of our two DSFGs highlight that such galaxies may not consistently reside in the dense cores of protoclusters. Instead, they may sometimes be located in the outskirts or in other regions of large-scale structures, though a larger sample is needed to establish whether this is a general trend. The lower number of companion galaxies in the SDP.17b environment could indicate that SDP.17b resides on the outskirts of a large-scale structure, potentially in an unbound region, consistent with a typical field. On the other hand, HELMS-55 likely resides closer, if not within, the core of a protocluster, given our results. 

\subsubsection{The Intrinsic Properties of Our DSFGs}

\begin{figure}
    \centering
    \includegraphics[width=1\linewidth]{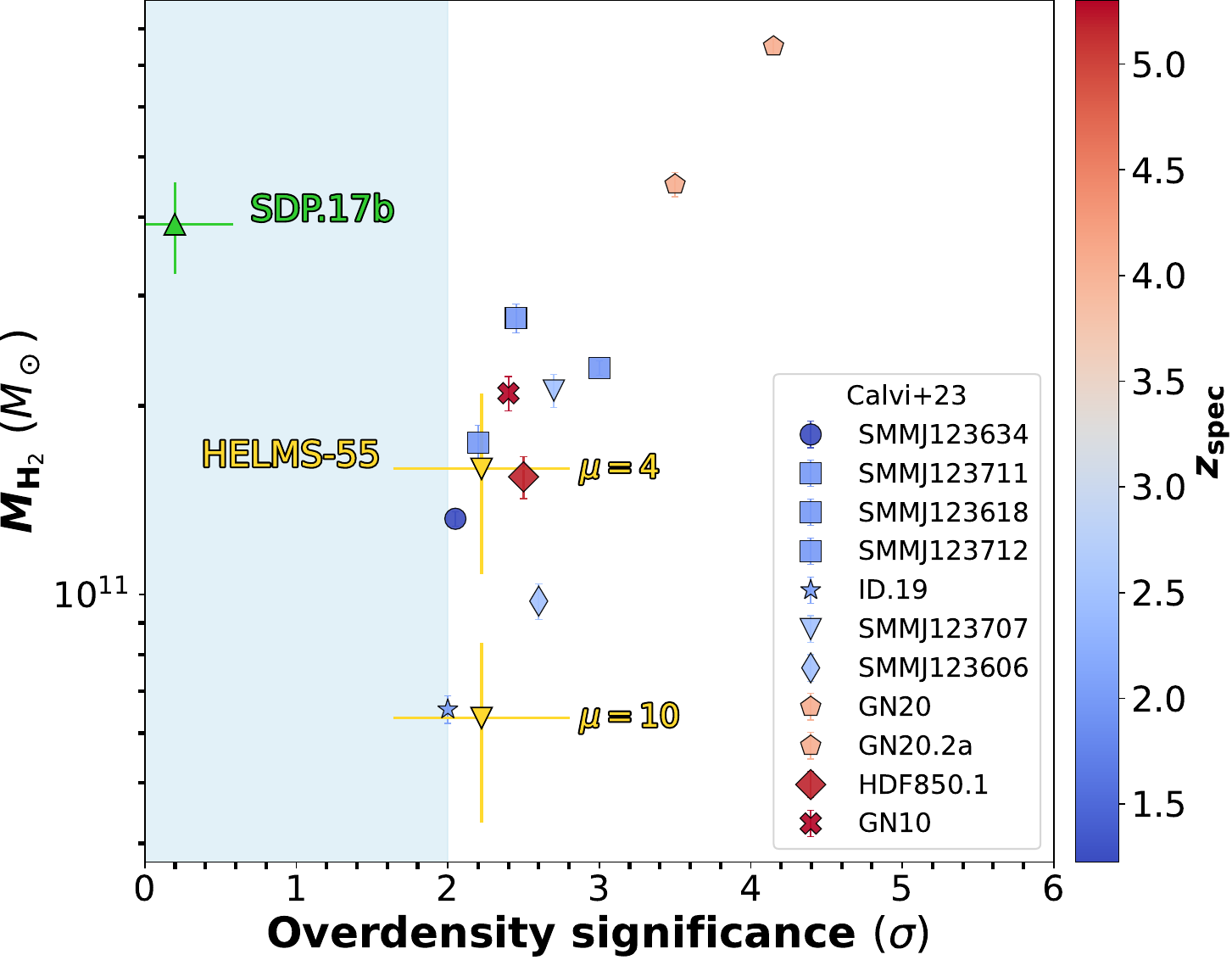}
    \caption{Molecular gas masses ($M_{\mathrm{H}_2}$) of our target DSFGs compared with the \citet{Calvi_2023} DSFGs sample, assuming $\alpha_\text{CO} =4.0\, M_\odot\,(K\,{\rm km\,s^{-1}\,pc^2})^{-1}$. SDP.17b is shown in green and HELMS-55 in yellow. For HELMS-55, we show estimates assuming different lensing magnifications ($\mu=4$ and $\mu=10$), since no lensing model currently exists in the literature. The shaded blue region marks the threshold below which overdensity significances are considered protocluster candidates. For the \citet{Calvi_2023} DSFGs, average overdensity significances were adopted; their colors correspond to their spectroscopic redshifts as indicated by the color bar, and marker shapes distinguish the individual DSFGs listed in the figure, where DSFGs near each other were grouped.}
    \label{fig:MH2_vs_overdensity_sig}
\end{figure}

Recent work has explored whether the molecular gas content of DSFGs correlates with environmental density. \citet{Calvi_2023} reported a positive correlation between $M_{\rm H_2}$ and overdensity significance, suggesting that DSFGs with larger gas reservoirs may preferentially reside in overdense regions. To test this proposed correlation, we derived molecular gas masses for SDP.17b and HELMS-55 from their observed CO fluxes.

Based on the CO flux measurements of $S\Delta v = 9.10$ Jy km s$^{-1}$ for SDP.17b (CO(4–3); \citealt{Omont_2013}) and $S\Delta v = 4.46$ Jy km s$^{-1}$ for HELMS-55 (CO(3–2); \citealt{Cox_2023}), we estimate molecular hydrogen masses of $(3.9 \pm 0.7) \times 10^{11}\, M_\odot$ and $(1.6 \pm 0.5) \times 10^{11}\, M_\odot$, respectively. For HELMS-55, we assumed a fiducial magnification factor of $\mu = 4$, since no lensing model currently exists. We adopted $\alpha_{\rm CO} = 4.0\, M_\odot\,(K\,{\rm km\,s^{-1}\,pc^2})^{-1}$ \citep{Berta_2023, Prajapati_2025}, and excitation ratios of $r_{41}=0.32$ and $r_{31}=0.60$ following \citet{Calvi_2023}. The positions of SDP.17b and HELMS-55 relative to the $M_{\rm H_2}$-overdensity-significance correlation differ: HELMS-55 may be consistent with the positive trend, whereas SDP.17b lies well below the protocluster threshold of $S=2\sigma$ despite a substantial gas reservoir, making it a possible outlier. The relation between $M_{\rm H_2}$ and overdensity significance, as defined by the \citet{Calvi_2023} DSFG sample, may provide a practical way to preselect DSFGs that are more likely to reside in overdense environments; however, our results remain inconclusive. A larger sample size and an accurate lensing model for HELMS-55 are required to robustly assess this correlation.

The intrinsic CO luminosities and linewidths of SDP.17b and HELMS-55 provide additional context for understanding how these systems compare to typical DSFGs. Even after correcting for lensing, both sources remain luminous with $L'_{\rm CO} \sim 4 \times10^{10}$ K km s$^{-1}$ pc$^2$, compared to a median $L'_{\rm CO} \sim 3 \times10^{10}$ K km s$^{-1}$ pc$^2$ from DSFGs in the literature \citep{Greve_2005, Bothwell_2013, Jin_2021}. Their narrow linewidths contrast with the broader median linewidths reported for DSFG samples ($\sim800$ km s$^{-1}$ in \citealt{Greve_2005}; $\sim 500$ km s$^{-1}$ in \citealt{Bothwell_2013}). Several of the \citealt{Bothwell_2013} sources are now known to reside in overdense environments \citep{Chapman_2009, Calvi_2023}, demonstrating that even within dense regions, DSFGs display diverse CO properties, and that SDP.17b and HELMS-55 fall into the narrow-FWHM but intrinsically luminous subset of this distribution.


\subsubsection{What Are the Implications of Our Results and the Literature?}

Our findings, together with results from the literature, suggest that DSFGs do not uniformly trace protocluster cores but instead occupy a diverse range of environments. In our sample, HELMS-55 resides in an overdensity consistent with a protocluster candidate; however, its companion galaxies are distributed between $\sim1-5$ cMpc in projected distance, while SDP.17b appears consistent with a field environment. This contrast suggests that DSFGs can be found both in the dense substructures of protoclusters and in the surrounding large-scale environments, similar to trends observed in other high-redshift systems.

Observational studies provide further evidence that DSFGs occupy a broad range of locations within protoclusters. Systems such as SSA22 \citep{Umehata_2017}, the Spiderweb protocluster \citep{Dannerbauer_2014, Zhang_2024, Perez_Martinez_2025_Spiderweb}, and MAMMOTH-1 \citep{Battaia_2018} contain DSFGs distributed from compact cores to extended filaments and infalling galaxy groups. The observed offsets between DSFGs and the highest-density regions span a wide range, from a few cMpc ($\sim2$–$3$ cMpc), consistent with the separation range found around HELMS-55, up to tens of cMpc ($\sim25$ cMpc). These patterns suggest that DSFGs may trace both early-stage, gas-rich filaments feeding cluster formation and later-stage, more evolved regions near the protocluster core.

Consistent with our target environments, narrowband photometry by \citet{Cornish_2024} also finds a diversity of DSFG environments, with sources located in protocluster regions, smaller proto-groups at $z_\text{spec}\sim2.3$, and even in underdense environments at $z_\text{spec}\sim3.3$. DSFGs have also been observed within protocluster cores. For example, \citet{Calvi_2023} report a median projected offset of $\sim2.1$ cMpc (MAD $\sim0.6$ cMpc) between DSFGs and the overdensity peak, with distances ranging from 0.3 to 3.6 cMpc (see Table 3 in \citealt{Calvi_2023}), a range consistent with the median projected separation of $\sim3.7$ cMpc found for companion galaxies around HELMS-55.

Our current sample of only two DSFGs is too small to draw a definitive conclusion of whether DSFGs are protocluster signposts. A larger sample is required to assess how DSFG properties such as intrinsic luminosity, line width, molecular gas content, and stellar mass may influence the environments that DSFGs trace. This is essential to determine whether particular types of DSFGs predominantly occupy specific regions within protoclusters or are more broadly distributed across environments. To verify this, other protocluster tracer populations need to be observed, such as Ly$\alpha$ emitters (LAEs), H$\alpha$ emitters, and rest-frame optical line emitters, which can more comprehensively map the underlying density distribution of galaxies.

\subsection{Why do Some DSFGs Reside Outside Protocluster Cores?}

A number of physical processes may explain why DSFGs can reside in the outskirts of protoclusters. \citet{Zhang_2024} find that ALMA-detected DSFGs are fueled by gas accretion along cosmic filaments, with intense star formation triggered by accretion shocks prior to their infall into the cluster core. Although \citet{Falgarone_2017} report evidence of shocks in SDP.17b, they interpret these shocks as being powered by starburst-driven galactic winds. Thus, while SDP.17b may reside in the outskirts of a protocluster, its infall status remains uncertain. 

More broadly, theoretical models suggest that roughly half of DSFGs with specific flux $S_{850} > 5$ mJy are triggered by mergers with mass ratios greater than 1:10, while the other half are powered by smooth accretion of cold gas from the intergalactic medium (IGM; \citealt{Dekel_2009, Casey_2014}). Large-scale structure may facilitate both mechanisms: star formation could be enhanced by inflows of gas along filaments of the cosmic web \citep{Umehata_2019}, or by an elevated likelihood of mergers in dynamically young, overdense regions that have not yet virialized into dense cluster cores \citep{Umehata_2017, Alberts_2022}. However, cold gas accretion is thought to be increasingly inefficient at $z \sim 2$ in halos above $\sim 10^{12}\,M_\odot$, where shock heating suppresses further inflow from the IGM \citep{Dekel_2006}. Observations of the Spiderweb galaxy, located at the center of the Spiderweb protocluster \citep{Perez_Martinez_2025_Spiderweb}, show that its molecular gas halo is dominated by recycled rather than pristine gas, reinforcing the idea that cold accretion is limited in dense cluster cores \citep{Emonts_2018}. Recent studies also suggest that galaxy mergers are more likely in dynamically cold regions of protoclusters with moderate velocity dispersions, and are suppressed in hot, virialized cores where high relative velocities prevent efficient merging \citep{Liu_2025}. Thus, there are indications that cold gas accretion and galaxy mergers may be more common in the outskirts of protoclusters than in their cores. This could help explain why DSFGs are not always found at the centers of protoclusters. The star formation rates and stellar masses of our DSFG companion galaxies may offer further insight into the physical conditions of the environments surrounding each DSFG.

\subsection{How do Galaxies Evolve Inside Our Target Environments?}

\begin{figure*}[t]
    \centering 
    \includegraphics[width=0.85\linewidth]{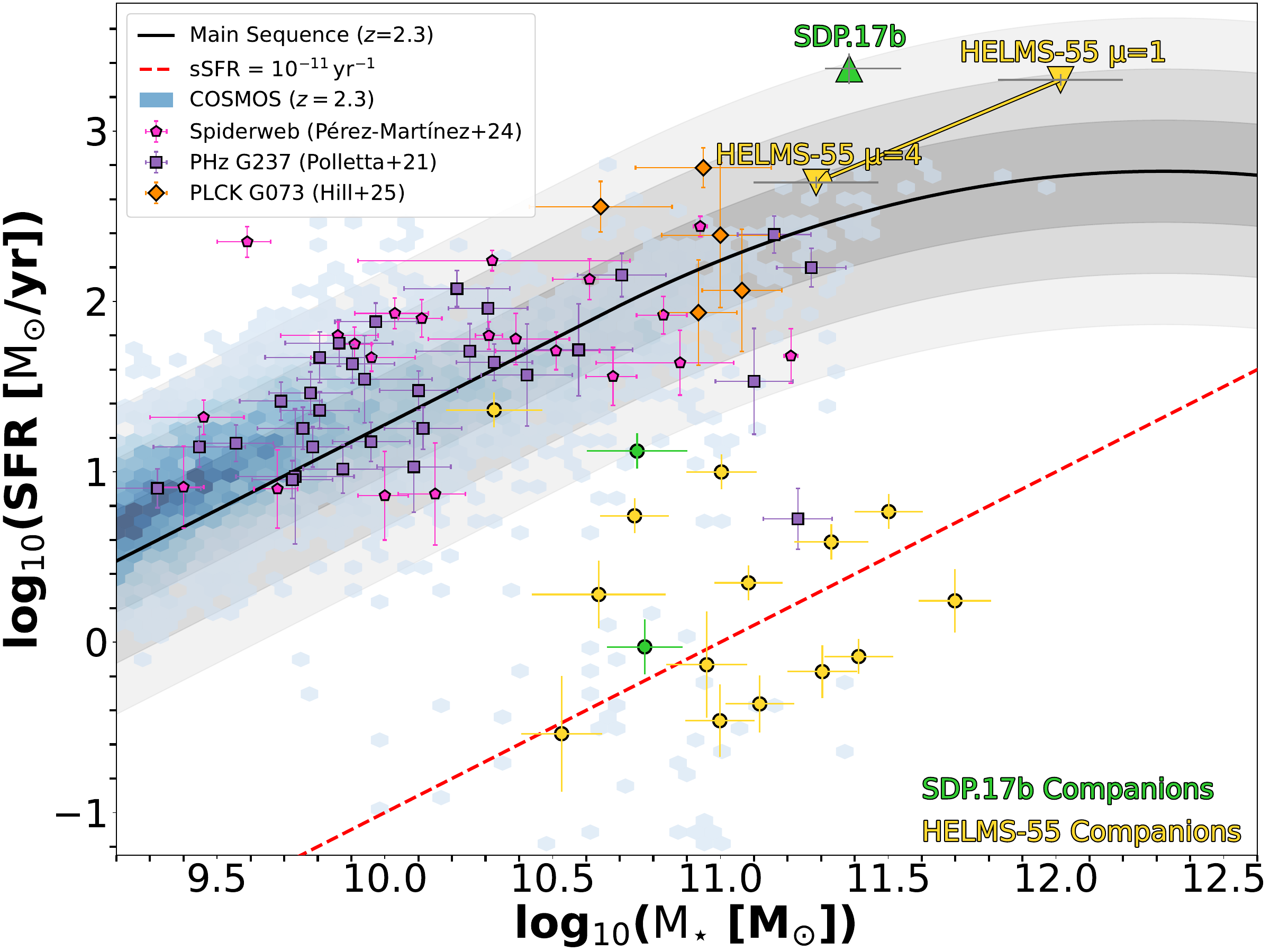}
    \caption{The \texttt{CIGALE} fit values for instantaneous SFR $[\text{M}_\odot \, \text{yr}^{-1}]$ and total stellar mass $[\text{M}_\odot]$ of the maximum likelihood companion galaxies (no magnitude cut applied) within $dz=0.2$ of the central DSFG $z_\text{spec}$, and the respective $1\sigma$ error bars. The main sequence at $z = 2.3$ from \citet{Schreiber_2015} is shown in black, with shaded regions indicating $\pm$0.3, 0.6, and 0.9 dex scatter in increasing transparency. We define quiescent galaxies as sources that fall below or are consistent with the red dotted line ($\text{sSFR} < 10^{-11} \, \text{yr}^{-1}$). The published SFR and $M_\star$ values of SDP.17b and HELMS-55 are also shown. For HELMS-55, we show a range of assumed magnification factors ($\mu=1$ and $\mu=4$) to account for the currently unconstrained lensing correction. For comparison, we plot published SFR and $M_\star$ values of known protoclusters from the literature: the Spiderweb protocluster \citep[$z=2.16$]{PMartinez_2024}, PHz G237.01+42.50 \citep[$z=2.16$]{Polletta_2021}, and PLCK G073.4-57.5 \citep[$z=2.3$]{Hill_2025}. Additionally, we plot field galaxies from the COSMOS survey \citep{COSMOS_2022} within $dz=0.2$ of $z=2.3$  as hexagonal density bins, where darker blue indicates higher source density.}
    \label{fig:sfr_vs_mstar}
\end{figure*}

The \texttt{CIGALE}-derived SFRs and stellar masses allow us to assess how the companion galaxies in the HELMS-55 environment compare to expectations for protocluster evolution. We present the \texttt{CIGALE} derived instantaneous SFRs and stellar masses for the maximum likelihood companion galaxies in each target environment, compared to the main sequence at $z \sim 2.3$ in Figure \ref{fig:sfr_vs_mstar}. These measurements also allow us to examine whether the properties of HELMS-55's companion galaxies are consistent with a transitional phase following the peak of star formation predicted in simulations such as the Manhattan Suite \citep{Rennehan_2024}, which reproduce elevated SFR densities and specific star formation rates (sSFRs) in protoclusters at $z \sim 4$. We include the SDP.17b environment for comparison, as a likely field, to help contextualize any environmental differences. The main sequence from \citet{Schreiber_2015} serves as a proxy for field SFRs at the average spectroscopic redshift of SDP.17b and HELMS-55 ($z_\text{spec} = 2.2942$). For reference, we also plot the published SFR and $M_\star$ values of the central DSFGs. SDP.17b has an SFR of $2300 \pm 500 \, \text{M}_\odot \, \text{yr}^{-1}$ and a stellar mass of ($24.2^{+8.6}_{-4.0}) \times 10^{10} \, M_\odot$ from \citet{Negrello_2014}, corrected for a gravitational lensing magnification of $\mu = 3.56^{+0.19}_{-0.17}$. No lensing model for HELMS-55 is currently available, so we present its estimated properties without a lensing correction: $\mu \text{SFR} = 2.0 \times 10^3 \pm 200 \, \text{M}_\odot \, \text{yr}^{-1}$, and $\mu^E M_\star = (10 \pm 4) \times 10^{11} \, M_\odot$, where $E = 1.21$ is derived from inversion of stellar mass scaling relations \citep{Berta_2023}. Throughout, we define quiescent galaxies as those with a sSFR $\leq 10^{-11} \, \text{yr}^{-1}$ (e.g., \citealt{Ilbert_2013, Florez_2020, Houston_2023}). Fourteen companion galaxies are plotted for the HELMS-55 environment rather than thirteen, because we include all companions without applying a magnitude cut. One galaxy is fainter than this magnitude cut, which is why Table \ref{tab:realizations_compared_to_COSMOS} lists 13 HELMS-55 maximum likelihood companion galaxies.

The maximum likelihood companion galaxies of both target environments lie primarily below the main sequence in Figure \ref{fig:sfr_vs_mstar}. About half of the SDP.17b companion galaxies, and one of the HELMS-55 companion galaxies, appear within 3 $\times$ the scatter of the main sequence (0.9 dex). However, the SDP.17b environment contains only two maximum likelihood companion galaxies, so this fraction is not statistically meaningful. Notably, approximately half of the HELMS-55 companion galaxies appear quiescent. This hints at the presence of environmentally quenched galaxies at cosmic noon, consistent with the scenario in which dense environments suppress star formation at $z<3$ \citep{Shimakawa2018b}. \citet{Kiyota_2024} report a similar excess of quiescent galaxies in protocluster candidates at $z \sim 2$, compared to field galaxies. These SFR and $M_\star$ estimates could indicate the HELMS-55 environment protocluster candidate is in the maturing phase of cluster evolution, where galaxies are expected to quench inside-out from the protocluster cores \citep{Shimakawa2018b}, assuming HELMS-55 traces a protocluster core. In contrast, the absence of quiescent galaxies in the SDP.17b environment could suggest that it lies in the outskirts of a forming protocluster, where environmental quenching is less efficient. This may also help explain why its measured overdensity is consistent with the field, as the region surrounding SDP.17b may still be unbound. Caution is warranted, as our photometry primarily samples blueward of the $1.6\mu$m stellar bump, making it less sensitive to older stellar populations. This limitation affects our ability to constrain dust attenuation and $M_\star$, introducing a systematic uncertainty in stellar mass estimates of $\sim$0.2–0.3 dex \citep{Kartheik_2025}.  Spectroscopic confirmation is essential for more robust redshift estimates and improved SFR and $M_\star$ measurements to confirm this trend. In addition, more NIR bands would help better constrain the SED shape, as young dusty galaxies tend to show steeper rest-frame UV slopes due to dust attenuation, while dust-free old galaxies typically feature prominent Balmer breaks.

For comparison, Figure \ref{fig:sfr_vs_mstar} includes known protoclusters from the literature alongside field galaxies from the COSMOS survey, with all published SFR and $M_\star$ values derived under broadly consistent assumptions with ours (e.g., a Chabrier IMF, solar metallicity, and the Calzetti dust attenuation law). COSMOS field galaxies were selected following the same redshift and quality cuts as in Section \ref{subsubsec:field_galaxies}, centered on the average spectroscopic redshift of SDP.17b and HELMS-55 ($z_\text{spec} = 2.2942$), with SFRs and stellar masses estimated using the SED-fitting software \texttt{LePHARE} \citep{LePHARE_2011}. The literature protoclusters plotted for comparison include the Spiderweb protocluster \citep[$z=2.16$]{PMartinez_2024}, PHz G237.01+42.50 \citep[$z=2.16$]{Polletta_2021}, and PLCK G073.4–57.5 \citep[$z=2.3$]{Hill_2025}. These are not intended to form an exhaustive or representative sample of all $z \sim 2$ protoclusters but rather to serve as illustrative examples that place our results in the context of known protoclusters.

Notably, the HELMS-55 protocluster candidate lies well below both the COSMOS field locus and the other literature protoclusters, suggesting a suppressed star-forming environment. In contrast, systems such as the SDP.17b environment, the Spiderweb, PHz G237.01+42.50, and PLCK G073.4-57.5 protoclusters are predominantly consistent with the main sequence and the COSMOS field population, indicating ongoing star formation. Previous studies have also reported both enhanced (e.g., \citealt{Shimakawa_2018,Monson_2021}) and suppressed (e.g., \citealt{Tran_2015}) star formation in protoclusters. This diversity suggests that, analogous to the variety of galaxy types, there may not be a single \lq\lq universal" protocluster environment: some appear to boost star formation, others maintain field-like activity, and some, like HELMS-55, may already be in a quenching phase, suppressing star formation.

Our findings provide a comparison point for theoretical models. Although cosmological simulations have made progress in tracing protocluster evolution (e.g., \citealt{Ata_2022, Bassini_2020, Remus_2023}), many continue to underestimate the SFRs observed in protocluster galaxies at $z > 2$ \citep{Lim_2021}, in part due to the lack of dust modeling. The F2 imaging field of view covers a radius of $\sim5.5$ cMpc around HELMS-55, likely probing regions beyond the core of a potential protocluster. The semi-analytic models of \citet{Chiang_2017} estimated that, when considering the total mass including dark matter, protoclusters have an average core radius of $\sim 0.7$ cMpc and a total-mass radius of $\sim8$ cMpc at $z\sim2.3$. These values suggest that the HELMS-55 environment lies within the expected extent of a protocluster, and that our observations likely sample beyond its dense core, where galaxies may already be quenching or transitioning out of peak star-forming phases.

Spectroscopic follow-up with F2 would be valuable for the SDP.17b and HELMS-55 environments. Spectra from such observations would allow us to confirm the redshifts of companion galaxies, enabling more accurate overdensity estimates for each target environment at $z \sim 2.3$. With H$\alpha$ emission and continuum measurements, we could directly probe the unobscured star formation rates \citep{Kennicutt_1998} and derive stellar masses of companion galaxies. These observations would help confirm whether SDP.17b and HELMS-55 are reside in protoclusters, and potentially reveal the effect of protocluster environments on the star formation activity of their members.

\section{Conclusions}\label{sec:conclusions}

This pilot study examined the environments of two lensed DSFGs at $z \sim 2.3$ to assess whether such luminous infrared galaxies can serve as reliable tracers of protoclusters. Our analysis is based on NIR $K_s$-band imaging from Gemini-South/F2, supplemented by optical HSC \textit{grizy} and NIR HST F110W and F160W imaging. We performed SED fitting to estimate photometric redshifts, star formation rates, and stellar masses for galaxies in the environment around each DSFG. Our main findings are:
\begin{itemize}
    \item SDP.17b exhibits a low overdensity, with a significance of $(0.2 \pm 0.4)\sigma$ relative to the number of galaxies expected from the COSMOS survey. This corresponds to an overdensity of $\delta = 0.1 \pm 0.2$, consistent with a typical field environment.
    \item HELMS-55 shows a more substantial overdensity, with a significance of $(2.2 \pm 0.6)\sigma$ relative to the COSMOS survey expectation. This corresponds to an overdensity of $\delta = 1.0 \pm 0.3$, making it a protocluster candidate. Based on lower-limit halo mass estimates, the HELMS-55 environment is expected to evolve into a present-day Fornax-type cluster. Approximately half of the HELMS-55 maximum-likelihood companion galaxies are quiescent, hinting at a potential link between $z \sim 2$ protocluster environments and quiescent populations.
    \item Although our results show contrasting environments for the two DSFGs, the small sample size precludes firm conclusions on whether luminous DSFGs consistently trace the most overdense regions.
    \item Our results are inconclusive with regard to the $M_{\rm H_2}$-overdensity-significance relation proposed by \citet{Calvi_2023}. HELMS-55 may be consistent with the relation (subject to lensing uncertainties); however, SDP.17b lies well outside it. A larger sample and an accurate lensing model for HELMS-55 are needed to robustly assess this correlation.
\end{itemize}

Our Gemini results demonstrate that wide-field surveys are essential for probing regions beyond the protocluster core, where the extended environments of DSFGs become apparent. There are substantial ongoing efforts in ground-based wide-field NIR imaging (e.g., the SuMIRe project; \citealt{Takada_2014}), as well as space missions such as \textit{Euclid} and \textit{Roman}, which are ideally suited to this goal. Future work should therefore focus not only on expanding sample sizes but also on extending follow-up observations to rest-frame NIR wavelengths redward of $2\mu$m (observed frame), enabling the detection of older and less massive member galaxies that may remain unseen in bluer bands. Such efforts will yield a more complete picture of the galaxy populations associated with DSFGs and their role as protocluster signposts.

\section{Acknowledgments}
The authors thank Flora Stanley, Roberto Neri, Alain Omont, and Stefano Berta for their contributions to the Gemini South Flamingos-2 $K_s$-band imaging proposals, which were essential to this project. ASWM thanks the organizers and participants of the \lq\lq From Fake News to Real Clusters: The Controversial Fate of High-z Galaxy Protoclusters" workshop for useful discussions. The authors thank the anonymous referee for constructive comments that improved the clarity and quality of the paper.

JB thanks the members of the UBC Extragalactic group for their support throughout this project. In particular, Anan Lu for insightful feedback and suggestions, and Laya Ghodsi and Lucas Kuhn for their helpful input and discussions. JB also thanks George Wang, Yunting Wang, and Ryley Hill for their valuable advice and many helpful conversations. JB is grateful to the Gemini Observatory research staff, Kathleen Labrie and Chris Simpson, for their patience and guidance in responding to countless Helpdesk inquiries. JB expresses heartfelt thanks to family and friends for their support, in particular, Iqbal Malhi, Michael Reeves, Jen Bhangal, Neel Soman, Michael Quinsey, Daniel Crook, Emma Klemets, Kian Jansepar, and Kevin Sohn. Oreo and Daisy deserve special recognition as post-dogs, whose contributions to stress relief were essential to the completion of this work.

JB and AWSM acknowledge the support of the Natural Sciences and Engineering Research Council of Canada (NSERC) through grant reference number RGPIN- 2021-03046. DD acknowledges support from the Polish National Agency for Academic Exchange (NAWA) grant No. BPN/BEK/2024/1/00029/DEC/1 and from the National Science Center (NCN) grant SONATA (UMO-2020/39/D/ST9/00720). HD acknowledges support from the Agencia Estatal de Investigaci\'on del Ministerio de Ciencia, Innovaci\'on y Universidades (MCIU/AEI) under grant (Construcci\'on de c\'umulos de galaxias en formaci\'on a trav\'es de la formaci\'on estelar oscurecida por el polvo) and the European Regional Development Fund (ERDF) with reference (PID2022-143243NB-I00/10.13039/501100011033).

Based on GS-2022B-Q-312, GS-2023B-Q-316, GS-2025A-Q-308 observations obtained at the international Gemini Observatory, a program of NSF NOIRLab, acquired through the Gemini Observatory Archive at NSF NOIRLab and processed using DRAGONS (Data Reduction for Astronomy from Gemini Observatory North and South; \citealt{DRAGONS_2023, Zendo_dragons}), which is managed by the Association of Universities for Research in Astronomy (AURA) under a cooperative agreement with the U.S. National Science Foundation on behalf of the Gemini Observatory partnership: the U.S. National Science Foundation (United States), National Research Council (Canada), Agencia Nacional de Investigaci\'{o}n y Desarrollo (Chile), Ministerio de Ciencia, Tecnolog\'{i}a e Innovaci\'{o}n (Argentina), Minist\'{e}rio da Ci\^{e}ncia, Tecnologia, Inova\c{c}\~{o}es e Comunica\c{c}\~{o}es (Brazil), and Korea Astronomy and Space Science Institute (Republic of Korea).



\facilities{Gemini South F2, HSC, HST}

All the HST data used in this paper can be found in MAST for the SDP.17b environment, \dataset[10.17909/djsv-d139]{http://dx.doi.org/10.17909/djsv-d139} and the HELMS-55 environment, \dataset[10.17909/y4cg-6t33]{http://dx.doi.org/10.17909/y4cg-6t33}.

\software{\texttt{SExtractor} \citep{SExtractor}, \texttt{EAZY-PY} \citep{Brammer_2008}, \texttt{CIGALE} \citep{Boquien_2019}}


\bibliography{sample631}
\bibliographystyle{aasjournal}

\appendix

\section{Appendix Information}

\subsection{Gemini South Observation Log}\label{subsec:observation_log}
\begin{table*}[ht]
    \centering
    \begin{tabular}{|c|c|c|c|c|c|} \hline 
         \textbf{Target environment}&\textbf{Exposure Time}  &\textbf{Exposure Time}& \textbf{Program ID}&\textbf{Observation}& \textbf{Usable Survey}\\
 & &\textbf{with Artifacts} & && \textbf{Area} \\
 & [s]& [s]& & &[$\text{arcmin}^2$]\\\hline  
         SDP.17b& 
       9587&6644& GS-2022B-Q-312&2022: Oct. 29-30  $|$ Nov. 6 $|$ Dec. 4, 6, 27& 31.46\\
 &  && &2023: Jan. 3, 4 10 &  \\ \hline
 SDP.17b& 9072& 0& GS-2025A-Q-308& 2025: April 9, 12, 13&33.12\\ \hline
 HELMS-55& 9782 &0& GS-2023B-Q-316&2023: Sept. 25, 27& 31.03\\  \hline \end{tabular}
    \caption{Observation log for the Gemini South Flamingos-2 $K_s$-band imaging. The exposure time given in column two refers to the total exposure time (including the data with artifacts for SDP.17b). The third column corresponds to the total exposure time of the data with the wave-like artifacts described in Section \ref{subsec:Ks_band_imaging}.}
    \label{tab:image_calibration}
\end{table*}

\subsection{Source Extraction and Photometry}\label{subsec:source_ext_and_phot}

\textbf{\texttt{SExtractor} parameters:}
\begin{table*}[ht]
    \centering
    \begin{tabular}{|c|c|c|}
    \hline
 \textbf{Parameter}&\textbf{Value}(s)&\textbf{Definition}\\ \hline 
         \texttt{DETECT\_MINAREA}&  6&Minimum number of pixels above threshold\\ 
         \texttt{DETECT\_THRESH}&  2.25&Detection threshold [Background RMS]\\
 \texttt{FILTER}& \texttt{Y}&Whether to apply the filter for detection\\
 \texttt{FILTER\_NAME}& (a)&Name of the file containing the filter\\ 
         \texttt{DEBLEND\_NTHRESH}&  32&Number of deblending sub-thresholds\\
 \texttt{DEBLEND\_MINCONT}& 0.0015&Minimum contrast parameter for deblending\\
 \texttt{CLEAN}& \texttt{Y}&Whether to clean spurious detections\\
 \texttt{CLEAN\_PARAM}& 1.0&Cleaning efficiency\\  
 \texttt{WEIGHT\_GAIN}& \texttt{N}&Whether to modulate \texttt{GAIN} with weights\\ 
 \texttt{PHOT\_APERTURES}& 3.0&Aperture diameter [pixels]\\
 \texttt{MAG\_ZEROPOINT}& (b)& Magnitude zeropoint [AB mag]\\
 \texttt{BACK\_SIZE}& 64&Background mesh size\\
 \texttt{BACK\_FILTERSIZE}& 3&Background filter size\\ \hline
    \end{tabular}
    \caption{The main SExtractor parameters used to perform source detection and photometry. Note that \texttt{Y} means \textit{Yes} and \texttt{N} means \textit{No}.
     (a) For the detection filters, we use a $9 \times 9$ and $7 \times 7$ pixel convolution kernels for SDP.17b and HELMS-55, respectively, based on the FWHMs of their $K_s$-band images.
     (b) Table \ref{tab:photometric_data} contains the AB magnitude zeropoint values used for the $K_s$-band and ancillary images.}
    \label{tab:main_SExtractor_params}
\end{table*}

The main \texttt{SExtractor} parameters used are shown in Table \ref{tab:main_SExtractor_params}, and we summarize the key settings used as follows. Following \cite{Estrada_2023} a model of the sky background was computed using a mesh size of $64 \times 64$ pixels and smoothed with a median filter of $3 \times 3$ pixels over the grid of meshes. Image pixels were then background-subtracted, and filtered using Gaussian point-spread functions (PSFs) convolutions. The aforementioned PSFs were chosen from \texttt{SExtractor} based on the FWHM of the $K_s$-band image. Sources were detected by flagging pixels with counts (ADUs) at least 2.25 times the background root-mean-square (RMS) noise or more. Sources were defined as objects with at least six contiguous pixels above the detection threshold. Following \cite{Postman_2012}, detections were segmented from the sky background and then deblended with a minimum contrast ratio of 0.0015 and 32 sub-threshold levels.

\newpage
\subsection{Companion Galaxies}\label{subsubsec:companion_galaxies_before_mag_cut}
\begin{table*}[ht]
    \centering
    \begin{tabular}{|c|c|c|c|} \hline 
         \textbf{Target environment}&  $dz$  &  $\overline{N}_\text{target, no mag cut}$ &\textbf{Maximum Likelihood}\\
 &   &  &\textbf{Companion Galaxies} (no mag cut)\\ \hline 
         &  0.1  &  3 $\pm$ 2&1\\ 
         SDP.17b&  0.2  &  6 $\pm$ 2&2\\ 
         &  0.3  &  9 $\pm$ 3&7\\\hline 
 & 0.1  & 8 $\pm$ 3&10\\ 
 HELMS-55& 0.2  & 16 $\pm$ 4&14\\ 
 & 0.3  & 25 $\pm$ 4&25\\ \hline
    \end{tabular}
    \caption{The average number of companion galaxies identified in both target environments, and the number of maximum likelihood companion galaxies, before a magnitude cut in the $K_s$-band was applied.}
    \label{tab:realizations_no_mag_cut_numbers}
\end{table*}

\begin{figure}[ht]
    \centering
    \includegraphics[width=1\linewidth]{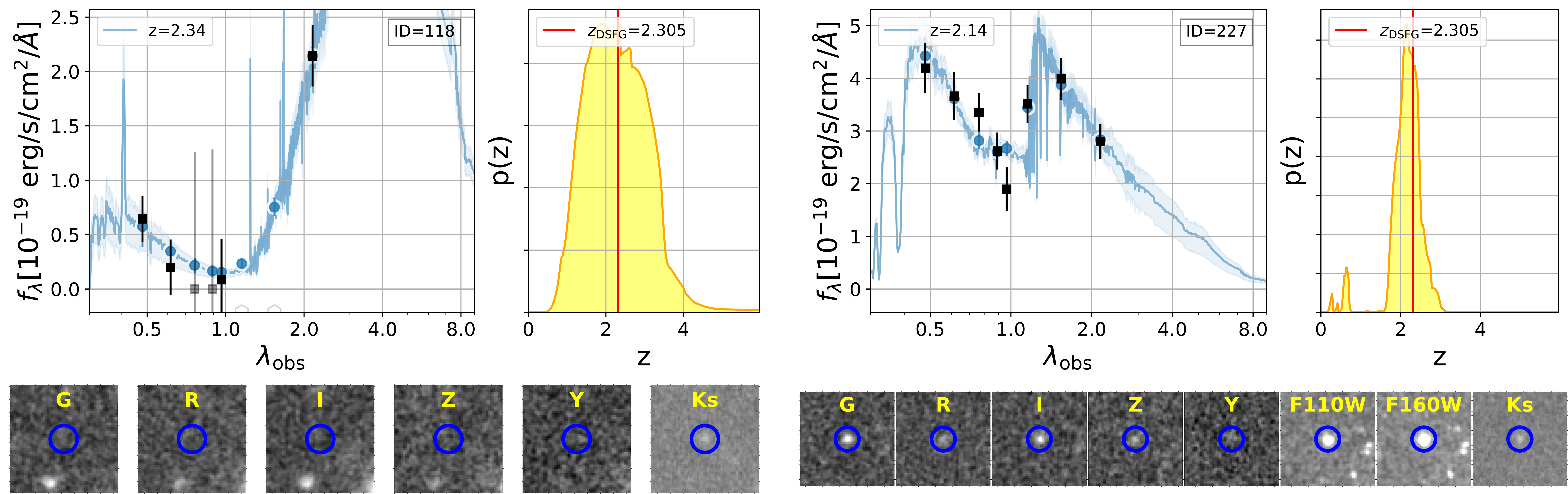}
    \caption{The description is the same as Figure \ref{fig:example_SEDs_zml}, except the sources are SDP.17b's maximum likelihood companion galaxies. SDP.17b's $z_\text{spec}$ is shown in red.}
    \label{fig:SDP17_companions}
\end{figure}

\begin{figure}[ht]
    \centering
    \includegraphics[width=1\linewidth]{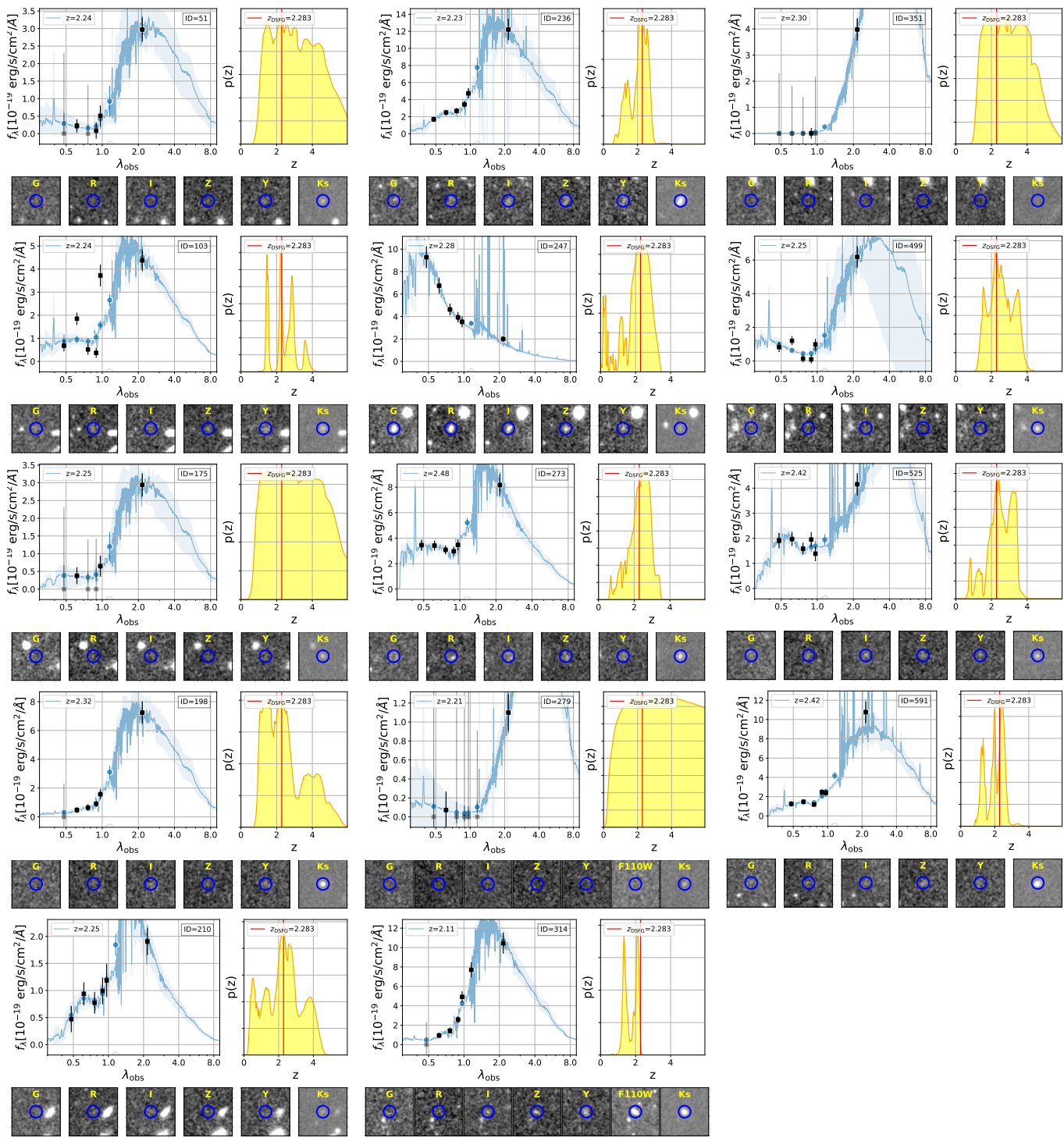}
    \caption{The description is the same as Figure \ref{fig:SDP17_companions}, except the sources are HELMS-55 maximum likelihood companion galaxies. HELMS-55's $z_\text{spec}$ is shown in red.}
    \label{fig:HELMS_55_companions}
\end{figure}

\clearpage
\subsection{Galaxy Properties Estimation}\label{subsec:galaxy_properties_estimation}
\textbf{\texttt{CIGALE} parameters:}
\begin{table*}[ht]
    \centering
    \begin{tabular}{|l|l|c|}
    \hline
          \textbf{Module}&\textbf{Parameter} &  \textbf{Value(s)}\\ \hline
          \texttt{sfhdelayed}&\texttt{tau\_main} ($\tau$) [Myr]&  250, 500, 750, 1000, 2000, 3000\\
  &\texttt{age\_main} ($t_0$) [Myr]& 500, 1000, 1500, 2000, 2500, 3000\\
 & \texttt{f\_burst}&0\\
          &\texttt{normalise}&  \texttt{True}\\ \hline
  \texttt{bc03}&\texttt{imf} & 1\\
  &\texttt{metallicity} &  0.02\\
  \hline
 & \texttt{E\_BV\_lines}&0.3\\
 & \texttt{E\_BV\_factor}&0.44\\
 & \texttt{uv\_bump\_wavelength} (nm)&217.5\\
 & \texttt{uv\_bump\_width} (nm)&0\\
 \texttt{dustatt\_modified\_starburst}& \texttt{powerlaw\_slope}&0\\
 & \texttt{Ext\_law\_emission\_lines}&1\\
 & \texttt{Rv}&3.1\\
 & filters&B\_B90 \& V\_B90 \& FUV\\\hline
    \end{tabular}
    \caption{The main \texttt{CIGALE} parameters. \texttt{sfhdelayed} sets the SFH, and \texttt{bc03} provides the stellar population properties. The default values of the dust attenuation module \texttt{\texttt{dustatt\_modified\_starburst}} were used. A \citet{Chabrier_2003} IMF is assumed. The \texttt{age\_main}, values were chosen assuming the DSFGs are maximally as old as the Big Bang (at maximum $\sim $3 Gyrs old at $z\sim 2.3$).}
    \label{tab:cigale_param}
\end{table*}
\end{document}